\documentclass[12pt,a4paper]{article}
\usepackage[latin1]{inputenc}
\usepackage{amsmath}
\usepackage{amsfonts}
\usepackage{amssymb}
\usepackage{graphicx}
\usepackage[affil-it]{authblk}
\usepackage{hyperref}
\usepackage{color}
\usepackage{caption}
\usepackage{verbatim}
\usepackage[normalem]{ulem}
\usepackage{subfigure}
\usepackage{slashed}

\newcommand{\phidag}{\phi^{\dagger}}
\newcommand{\vdag[1]}{V^{\dagger}_{#1}}
\newcommand{\Vdag[1]}{V^{#1\dagger}}
\newcommand{\Lbar}{\overline{L}}

\newcommand{\ld[1]}{\lambda_{#1}}
\newcommand{\Mv[1]}{M_{V^{#1}}}
\newcommand{\Mvc}{M_{V^{\pm}}}
\newcommand{\lamL}{\lambda_L}
\newcommand{\RD}{\Omega_{\text{DM}}h^2}

\usepackage{tikz}
\usetikzlibrary{arrows,shapes}
\usetikzlibrary{trees}
\usetikzlibrary{matrix,arrows} 				% For commutative diagram
\usetikzlibrary{positioning}				% For "above of=" commands
\usetikzlibrary{calc,through}				% For coordinates
\usetikzlibrary{decorations.pathreplacing}  % For curly braces
\usepackage{pgffor}							% For repeating patterns

\usetikzlibrary{decorations.pathmorphing}	% For Feynman Diagrams
\usetikzlibrary{decorations.markings}
\tikzset{
    vector/.style={decorate, decoration={snake}, draw},
	provector/.style={decorate, decoration={snake,amplitude=2.5pt}, draw},
	antivector/.style={decorate, decoration={snake,amplitude=-2.5pt}, draw},
    fermion/.style={draw=black, postaction={decorate},
        decoration={markings,mark=at position .55 with {\arrow[draw=black]{>}}}},
    fermionbar/.style={draw=black, postaction={decorate},
        decoration={markings,mark=at position .55 with {\arrow[draw=black]{<}}}},
    fermionnoarrow/.style={draw=black},
    gluon/.style={decorate, draw=black,
        decoration={coil,amplitude=4pt, segment length=5pt}},
    scalar/.style={dashed,draw=black, postaction={decorate},
        decoration={markings,mark=at position .55 with {\arrow[draw=black]{>}}}},
    scalarbar/.style={dashed,draw=black, postaction={decorate},
        decoration={markings,mark=at position .55 with {\arrow[draw=black]{<}}}},
    scalarnoarrow/.style={dashed,draw=black},
    electron/.style={draw=black, postaction={decorate},
        decoration={markings,mark=at position .55 with {\arrow[draw=black]{>}}}},
	bigvector/.style={decorate, decoration={snake,amplitude=4pt}, draw},
}\usetikzlibrary{decorations.markings}

\tikzstyle{block} = [draw, rectangle, 
    minimum height=3em, minimum width=6em]

\title{Dark Matter from a Vector Field in the Fundamental Representation of $SU(2)_L$}

\author[1,2]{Bastian D\'iaz Sáez}
\author[1,3]{Felipe Rojas-Abatte}
\author[1,2]{Alfonso R. Zerwekh}

\affil[1]{\normalsize Departamento de F\'isica, Universidad T\'ecnica Federico Santa Mar\'ia, Valpara\'iso, Chile} 
\affil[2]{\normalsize Centro Cient\'ifico-Tecnol\'ogico de Valpara\'iso, Casilla 110-V, Valparaíso, Chile} 
\affil[3]{\normalsize University of Southampton, Southampton, United Kingdom}

\date{}

\begin{document}
\maketitle

\begin{abstract}
  We explore an extension to the Standard Model which incorporates a vector field in the fundamental representation of $SU(2)_L$ as the only non-standard degree of freedom. This kind of field may appear in different scenarios such as Compositness, Gauge-Higgs unification and extradimensional scenarios. We study the model in which a $Z_2$ symmetry is manifiest, making the neutral CP-even component of the new vector field a vectorial dark matter candidate. We constraint the parameter space through LEP and LHC data, as well as from current dark matter searches. Additionally, comment on the implications of perturbative unitarity are presented. We find that the model is highly constrained but a small region of the parameter space can provide a viable DM candidate. On the other hand, unitarity demands an UV completion at an scale below 10 TeV. Finally we contrast our predictions on mono-$jet$, -$Z$, -Higgs production with the ones obtained in the inert Two Higgs Doublet Model.

  % Due to the similarities with the inert Two Higgs Doublet Model, we compare some physical predictions in both model.
\end{abstract}

\section{Introduction}
It is generally acknowledged that the Standard Model (SM) is incomplete, despite its impressive and unexpected phenomenological success. Its lack of a Dark Matter (DM) candidate and the impossibility of naturally generating within its framework tiny but non-vanishing neutrino masses are usually among the reasons invoked to illustrate such an incompleteness. Additionally, and maybe more dramatically, even the dynamical origin of the electroweak scale is not completely understood in within the SM. This has motivated, over the years, the construction of many extensions of SM. However, the very precise measurement made at LEP  during 1990's already taught us that the New Physics has to be subtle, making the construction of consistent and complete New Physics models, a formidable task. Of course, the lack of evidence of any kind of non-standard phenomena at the LHC has put stronger constrains and has made the labor of model-builders even more difficult.  

Under these circumstances, it seems wise to take a less ambitious approach. In the search of some dark matter signal, both effective field theories (EFT) (see e.g. \cite{Beltran:2010ww, Bai:2010hh, Fitzpatrick:2012ix, Busoni:2013lha, Busoni:2014sya, Belyaev:2016pxe, PhysRevD.85.056011, PhysRevD.84.095013}) and simplified models frameworks (see e.g. \cite{Buckley:2014fba, Buchmueller:2013dya, Buchmueller:2014yoa, Abdallah:2015ter, Abdallah:2014hon}) have been used as a guide of search. In the latter framework, both scalar and fermion dark matter has been the most explored line by their simplicity (for a classification under SM quantum numbers see \cite{Cirelli:2005uq}). However, it has been shown that vector bosons may perfectly play the role of dark matter, most of them motivated from hidden gauge sectors \cite{Hambye:2008bq, DiazCruz:2010dc, Kanemura:2010sh, Djouadi:2011aa, Lebedev:2011iq, Farzan:2012hh, Baek:2012se, Yu:2014pra, Gross:2015cwa, DiChiara:2015bua}, extra large dimensions \cite{Servant:2002aq}, little Higgs model \cite{Birkedal:2006fz} and from a linear sigma model \cite{Abe:2012hb}. Recently, the neutral component of an electroweak vector multiplet has been shown to be a good dark matter candidate, such as multiplets transforming in the adjoint representation \cite{Belyaev:2018xpf}, and in the fundamental one in the context of 331 models \cite{Mizukoshi:2010ky, Dong:2017zxo} and in Gauge-Higgs unification framework \cite{Maru:2018ocf}. 

In this work, taking an agnostic UV-completion approach, we consider a simplified model which takes an electroweak vector multiplet transforming in the fundamental representation of $SU(2)_L$ with hypercharge 1/2, leading naturally a dark matter candidate due to an accidental $Z_2$ symmetry. We introduce what we call the Dark Vector Doublet Model (DVDM), and we contrast this possibility with theoretical and experimental constrains. Interestingly, the new fields couples to the SM bosons ($Z$, $W^\pm$, photon and the Higgs boson) but, as we will see, not to the SM fermions if we only consider up to renormalizable operators. We constaint the model through experimental data such as LEP, LHC and dark matter probes. We show that our vector dark matter field can account for the total observed dark matter abundance for masses satisfying $\gtrsim 800$ GeV.
%contrasting other electroweak vector dark matter models (e.g.\cite{Mizukoshi:2010ky,Dong:2017zxo}), where the dark vector only can explain a tiny fraction of the budget.

Finally, DM cross section and missing transverse energy at the LHC are observables capables to discriminate among different signals. In view of the similarities between our model and the well known Inert Two Higgs Doublet Model (i2HDM)\cite{PhysRevD.18.2574}\cite{Barbieri:2006dq}\cite{Goudelis:2013uca}\cite{Arhrib:2013ela}\cite{Ilnicka:2015jba}\cite{Belyaev:2016lok}, we compare the cross section and missing energy distribution shapes for mono-jet, -$Z$ and -Higgs signals in both models.
% , which the latter contains four new scalars, but the couplings among them and the SM are very similar to the VDDMM. 

The paper is organized in the following way. In section \ref{lagrangian}
 we describe our model and we specify under what conditions a dark matter candidate appears.  The space parameter is discussed in section \ref{DMM}
% the full Lagrangian of the model is presented, then a shorted version of it compatible with a DM candidate and a discusion of the parameter space of the VDDMM is shown in section (\ref{DMM}).
We present as well all the relevant theoretical and experimental constraints we used to study the model in section \ref{constraints}. Next, in section \ref{DM_pheno} a deep description of the DM phenomenology of the model with a full scan of the parameter space is shown accompanied with a brief discussion of DM searches at LHC. As a complementary analysis, in section \ref{pu} we discuss perturbative unitarity. Finally we present our conclusions in section \ref{Conclusions}. In the appendix \ref{app:A} we show more details about the discussion of perturbative unitarity.

\section{The Lagrangian} \label{lagrangian}
As we announced, we extend the SM by introducing a new set of vector fields in a single-representation of the standard gauge group $SU(3)_c\times SU(2)_L\times U(1)_Y$:
\begin{equation}
 V_\mu = \begin{pmatrix}  V^{+}_\mu \\ V^{0}_\mu \end{pmatrix} = \begin{pmatrix} V^{+}_\mu \\  \frac{V^{1}_\mu + i V^{2}_\mu}{\sqrt{2}}, \end{pmatrix}
\end{equation}
transforming as $\left(\textbf{1},\textbf{2},1/2\right)$% \footnote{For an explicit UV-realization of this kind of vectors see \cite{Chizhov:2009fc}. They show the appearence of $SU(2)_L$ massive vector doublet as the spontaneously breaking of a $U(3)_W$ gauge symmetry to the $G_{SM}$ by some new scalar sector.}
. Notice that we are assuming that $V_\mu$ has the same weak-isospin and hypercharge than the Higgs doublet ($\phi$). In other words, we are assuming that $V^{1}_\mu$ and $V^{2}_\mu$ are neutral states. The most general Lagrangian containing this new vectors with operators up to dimension four is:
\begin{eqnarray} \label{eq:fl}
{\cal L} &=& -\frac{1}{2}\left(D_{\mu} V_{\nu} - D_{\nu} V_{\mu} \right)^{\dagger} \left(D^{\mu} V^{\nu} - D^{\nu} V^{\mu} \right) +  M_V^2 \vdag[\mu] V^{\mu} \nonumber \\
& + &\ld[2](\phidag \phi) (\vdag[\mu]V^{\mu}) + \ld[3](\phidag V_{\mu})(\Vdag[\mu]\phi)  + \ld[4](\phidag V_{\mu})(\phidag V^{\mu}) \nonumber \\ 
\nonumber &+&  \alpha_1 \phidag D_{\mu} V^{\mu} + \alpha_2 (\vdag[\mu] V^{\mu}) (\vdag[\nu] V^{\nu}) + \alpha_3 (\vdag[\mu] V^{\nu}) (\vdag[\nu] V^{\mu}) \nonumber \\
 &+& ig\kappa_1V_\mu^\dagger W^{\mu\nu} V_\nu + i\frac{g'}{2}\kappa_2 V_\mu^\dagger B^{\mu\nu} V_\nu  + h.c.
\end{eqnarray}
where
% $D_\mu$ is the same SM covariant derivative for the Higgs field, $\phi$,
$B^{\mu\nu}$ is the abelian $U(1)_Y$ field strenght, and $W^{\mu\nu}=W^{\mu\nu a}\frac{\tau^a}{2}$ is the non-abelian $SU(2)_L$ field strenght. In principle, all the free parameters, $\lambda_i$, $\alpha_i$ for $i=1,2,3$ may be complex. The parameters $\kappa_1$ and $\kappa_2$ are analogous the the well-known anomalous couplings in the context of vector leptoquark models.

Interestingly, due to the symmetries of the model, it is not possible to couple the new vector boson to the standard fermions with renormalizable operators. For example, let us suppose a Lorentz invariant Yukawa-like coupling between SM first generation of leptons and the vector doublet. Then, consider the following vector and axial vector couplings, 
\begin{eqnarray}\label{eq:fercoup}
 \mathcal{L} &\supset &  \Lbar \gamma^{\mu} \left(g^V  - g^A \gamma^5\right) e_R V_{\mu}
\end{eqnarray}
where $g^V$ and $g^A$ are unknown coupling constants. Considering the chirality projectors $P_{L,R}$, the Lagranian (\ref{eq:fercoup}) may be rewritten as
\begin{eqnarray*}
\mathcal{L} &=& \begin{pmatrix} \overline{\nu}_e & \overline{e} 
\end{pmatrix}
P_R \gamma^{\mu}\left(g^V  - g^A \gamma^5\right) P_R e V_{\mu}\\
&=& \begin{pmatrix} \overline{\nu}_e & \overline{e} 
\end{pmatrix}
P_R\gamma^{\mu} \left(g^V - g^A \gamma^5\right) P_LP_R e  V_{\mu} \\
&=& 0 
\end{eqnarray*}
where in the second line we have used the property $\{\gamma^\mu , \gamma^5\} = 0$, and in the last line we have used that $P_LP_R = 0$. This fact can be extrapolated straighforwardly to all SM fermions. 

On the other hand, the model allows a dimension three operator which is the only one linear in $V_\mu$. In principle, this term would introduce a mixing between the SM gauge bosons and the new vector states. However, it is possible to set up its corresponding coupling constant ($\alpha_1$ in (\ref{eq:fl})) to zero because an accidental $Z_2$ symmetry appears in the Lagrangian. Due to the new symmetry this choice
is technically natural in the sense of t'Hooft.
% in this work we refrain ourselves to the study in which this term is not present, because of the presense of an explicit $Z_2$ symmetry.
Therefore, in this limit and at the renormalizable level, the new vector sector only communicates to the SM through the electroweak gauge bosons, the photon and SM-Higgs boson. As a consequence, the flavour sector is untouch at tree level.

% Exotic vector-like fermions may be introduced in order to get a bridge between the SM fermions and the new vectors \cite{Chizhov:2009fc}. However, although vector-like partners are well motivated in the literature, we refrain ourselves from this possibility in favor of minimality.  

Finally, the terms in the last line of (\ref{eq:fl}) are allowed by the symmetry. However the value of their coupling constant ( $\kappa_1$ and $ \kappa_2$) are not fixed by the symmetries.  In this paper,  we work in the simplified case where $\kappa_1 = \kappa_2 = 1$. This choice is consistent with the hypercharge assigned to $V_{\mu}$ and agrees with what happen  in vector leptoquarks models, where the ultra-violet gauge completion and unitarity arguments fixes the values of those parameters to one \cite{Hewett:1993ks}. In other words, if we allow for values different to one, there appear the coupling among the photon $A_\mu$ and the two neutral vector $V^1$ and $V^2$, implying the latter fields now get an electric charge.
% , and this scenario deviate of our purpose of the present work. 
         
\section{Dark Matter Candidate} \label{DMM}
As we explained above, in the limit when $\alpha_1$ vanish, the model acquires an additional $Z_2$ discrete symmetry
% \footnote{This feature is not a benchmark point, but according to t'Hooft naturalness argument, it means that $\alpha_1$ could be though naturally having a value very close to zero.}
allowing the stability of the lightest odd particle (LOP). If the LOP happens to be a neutral component of $V_{\mu}$ (as it must be for cosmological reasons) then it constitutes a good DM candidate.
In this case, the Lagrangian (\ref{eq:fl}) reduces to:

\begin{eqnarray} \label{eq:lag2}
{\cal L} &=& -\frac{1}{2}\left(D_{\mu} V_{\nu} - D_{\nu} V_{\mu} \right)^{\dagger} \left(D^{\mu} V^{\nu} - D^{\nu} V^{\mu} \right) + M_V^2 \vdag[\mu] V^{\mu} \nonumber \\
&-& \alpha_2 (\vdag[\mu] V^{\mu}) (\vdag[\nu] V^{\nu}) - \alpha_3 (\vdag[\mu] V^{\nu}) (\vdag[\nu] V^{\mu}) - \lambda_2(\phidag \phi) (\vdag[\mu]V^{\mu}) \nonumber \\
&-& \lambda_3(\phidag V_{\mu})(\Vdag[\mu]\phi) - \frac{\lambda_4}{2} \left[(\phidag V_{\mu})(\phidag V^{\mu}) + (\Vdag[\mu]\phi)(\vdag[\mu]\phi)\right] \nonumber \\
 &+& i\frac{g'}{2} V_\mu^\dagger B^{\mu\nu} V_\nu + igV_\mu^\dagger W^{\mu\nu} V_\nu .  \label{DM-Lagrangian}
\end{eqnarray}

Curiously, this Lagrangian is rather similar to the i2HDM\cite{Barbieri:2006dq}\cite{LopezHonorez:2006gr}\cite{Belyaev:2016lok} where the extra scalar doublet is replaced by the new vector doublet.

The Lagrangian \ref{DM-Lagrangian} contain six free parameters\footnote{We assume that all the free parameters are real, otherwise, the new vector sector may introduce CP-violation sources. In this work we do not deal with that interesting possibility.} which we labelled as $\ld[2], \ld[3], \ld[4]$ for quartic coupling involving interactions between SM-Higgs field and the new vector field, a mass term $M_V$, and $\alpha_2, \alpha_3$ for quartic couplings of pure interactions among the vector fields. These latter self-interacting terms are not relevant for the experimental constraints and dark matter phenomenology done in this paper, therefore from now on we will not consider them, However, self-interacting particle dark matter can be relevant in related fields such as astrophysical structures \cite{Tulin:2017ara}.

After the electroweak Symmetry Breaking, the tree level mass spectrum of the new sector is
\begin{eqnarray}
M^2_{V^{\pm}} &=& \frac{1}{2}\left[2M_V^2 - v^2 \lambda_2 \right], \label{MassVc}\\
M^2_{V^1} &=& \frac{1}{2}\left[2M_V^2 - v^2(\lambda_2 + \lambda_3 + \lambda_4)\right], \label{MassV1}\\
M^2_{V^2} &=& \frac{1}{2}\left[2M_V^2 - v^2(\lambda_2 + \lambda_3 - \lambda_4)\right],  \label{MassV2}
\end{eqnarray}
% i.e., two charged degenerate states $V^{\pm}$ and two neutral $V^1$ and $V^2$. 
The term proportional to $\lambda_4$ makes the splitting between the physical masses of the two neutral states. For phenomenological proposes we will work in a different base of free parameters
% \footnote{{\color{red}{It should be noted that in some cases it will convenient to complement the analysis with the old base (e.g. in the Higgs diphoton decay subsection).}}}
\begin{equation}
\Mv[1], \quad \Mv[2], \quad \Mvc, \quad \lamL, \label{parameters}
\end{equation} 
where $\lamL=\ld[2]+\ld[3]+\ld[4]$ is, as we will see, the effective coupling controlling  the interaction between the SM Higgs and $V^1$.
% , which will be of very important in the phenomenological study, as will see later
%Fig.~(\ref{DM-vertex}).
%\begin{figure}[ht]
%\centering
%\begin{tikzpicture}[line width=1.0 pt, scale=1]
%\begin{scope}[shift={(0,0)}]
%	\draw[vector](-1,1) -- (0,0);
%	\draw[vector](0,0) -- (-1,-1);
%	\draw[scalar](0,0) -- (1.5,0);
%	\node at (-1.5,1) {$V_{\mu}^1$};
%	\node at (-1.5,-1) {$V_{\nu}^1$};
%	\node at (0.7,0.3) {$H$};
%	\node at (1.8,0) {:};
%	\node at (5,0) {$2\frac{M_W \sin{\theta_W}}{e}g^{\mu\nu}\ld[L]$};
%\end{scope}
%\end{tikzpicture}
%\caption{\footnotesize Leading Feynman diagram coupling two Dark Matter particles to the Higgs Boson.} \label{DM-vertex}
%\end{figure}
%\newline
It is convenient to write  the quartic coupling and the mass parameter as a function of the new free parameters
\begin{eqnarray} \label{eq:lam2}
\nonumber \ld[2] = \ld[L] + 2\frac{\left(\Mv[1]^2-\Mvc^2\right)}{v^2},  &\qquad&  \ld[3] = \frac{2\Mvc^2 - \Mv[1]^2 - \Mv[2]^2}{v^2}, \\
\ld[4] = \frac{\Mv[2]^2 - \Mv[1]^2}{v^2}, &\qquad& M_V^2 = \Mv[1]^2 + \frac{v^2\ld[L]}{2}. \label{lambda-couplings}
\end{eqnarray}
For future convenience, it will be useful to introduce
\begin{eqnarray}\label{eq:lamR}
 \lambda_R \equiv \ld[2]+\ld[3]-\ld[4] = \lambda_L + \frac{2\left(M_{V^2}^2 - M_{V^1}^2\right)}{v^2},
\end{eqnarray}
which is not a new free parameter, but it is the effective coupling constant which governs the $HV^2V^2$ interaction.
% as well as the quartic coupling of $V^1$ to the longitudinal $Z$ bosons $V^1V^1Z_LZ_L$. 

It is important to mention that because the new vector field have the same quantum numbers than the SM-Higgs field, the two neutral vectors have opposite CP-parities. However
% , from an experimental point of view it is impossible to distinguish between them. Besides,
 we can switch their parity just making a change of bases $V_{\mu}\rightarrow iV_{\mu}$ and then re-label each field as $V^1_{\mu} \rightarrow V^2_{\mu}$ and $V^2_{\mu} \rightarrow V^1_{\mu}$ and still obtaining the same phenomenology. Therefore, without loose of generality, we will choose $V^1_{\mu}$ as the LOP turning it into our Dark Matter candidate. Following the same line, to make sure that $V^1_{\mu}$ is the lightest state of the new sector, we can find some restrictions that the quartic couplings must follow to satisfy this condition. Considering this we can stress that
\begin{eqnarray}
\nonumber \Mv[2]^2 - \Mv[1]^2 > 0  \qquad & \Rightarrow &\ld[4] > 0, \\
\Mvc^2 - \Mv[1]^2 > 0 \qquad & \Rightarrow &\ld[3] + \ld[4] > 0 .
\end{eqnarray}
In order to have a weakly interacting model, we set that all the couplings parameters must to satisfy
\begin{equation}
|\lambda_i| < 4 \pi  \quad \wedge \quad |\alpha_j| < 4 \pi \qquad (i=2,3,4;\quad j=2,3). \label{pert}
\end{equation} 

%\subsubsection*{Model implementation}

We implemented this model using the \texttt{LanHEP}\cite{Semenov:2010qt} package  and we used \texttt{CalcHEP}\cite{Belyaev:2012qa} and \texttt{micrOMEGAs}\cite{Belanger:2013oya, Belanger:2006is, Belanger:2010gh} for collider and DM phenomenology calculations, respectively.

% for automatic Feynman rules derivation. We included effective vertex $Hgg$ and $H\gamma\gamma$ and we performed a cross-check of the gauge invariance implementation calculating several 2$\rightarrow$2 processes in both gauges (Unitary and 'tHooft - Feynman gauge) using \texttt{CalcHEP}\cite{Belyaev:2012qa} program. The Dark Matter penomenology was made with the \texttt{micrOMEGAs}\cite{Belanger:2013oya, Belanger:2006is, Belanger:2010gh} package. This program solves the Boltzmann equation numerically and calculate all of the relevant annihilation cross sections involved in the process using the \texttt{CalcHEP} program. \texttt{micrOMEGAs} consider as well the followign effects:
% \begin{itemize}
% \item The case when $\Mv[1]<M_W,M_Z$ taking into account the annihilation into 3-body final state from $V V^*$ or 4-body final state from $V^*V^*$ ($V = W^{\pm} , Z$). 
% \item The (co)annihilation effects when the split between the DM mass and the other
% particles is small. We took into account the $V^1 - V^2$ , $V^1 - V^{\pm}$ and $V^2 - V^{\pm}$ cases.
% \item The spin-independent cross section of DM scattering off the proton. 
% \end{itemize}
% The model was implementes using the independent parameter mentioned in (\ref{parameters}).

\section{Constraints from LEP, LHC, DM relic density and Direct Detection experiments}\label{constraints}
%%%%%%%%%%%%%%%%%%%%%%%
\subsection{LEP limits}
%%%%%%%%%%%%%%%%%%%%%%%
Considering that the coupling between the SM gauge bosons and the dark sector is fixed by gauge invariance, the only way to avoid deviations from precise LEP-I constraints on $W$ and $Z$ widths \cite{Cao:2007rm}\cite{Gustafsson:2007pc} is to demand that the channels $Z\rightarrow V^1V^2,V^+V^-$ and $W^\pm\rightarrow V^1V^\pm, V^2V^\pm$ are kinematically not open. This leads to the following conditions on the masses
%put lower bounds on their masses. For example, if $\Mv[1] + \Mv[2] < M_Z$, the $Z$-boson can decay into $V^1V^2$ pair with the subsecuently decay $V^2 \rightarrow V^1+\overline{f}f$, where $f$ can be either a SM quark or a lepton (see Fig.({\ref{LEPII}})). These kind of processes are very constrained by LEP-I precision measurements\cite{Kobel:315298}, therefore, in order to avoid these signals, we impose the following constraints on the masses:
\begin{eqnarray}
\nonumber \Mv[1] + \Mvc > M_{W^{\pm}}, &\qquad& \Mv[2] + \Mvc > M_{W^{\pm}}, \\ 
	  \Mv[1] + \Mv[2] > M_Z, &\qquad& 2\Mvc > M_Z.  \label{LEPI}
\end{eqnarray}
On the other hand, bounds on supersymmetric particles searches at LEP has been very useful to constraint other models beyond SM. In particular, LEP-II limits on neutralinos and charginos has been used to constraint the inert doublet model (i2HDM) \cite{PhysRevD.79.035013, Pierce:2007ut}. Although there are some differences in the number of Feynman diagrams and the spin involved in the processes, the kinematical efficiencies among the two result to be quite similar, allowing to recast the experimental bounds.

\begin{figure}[htb]
\centering
{\includegraphics[width=0.50\textwidth]{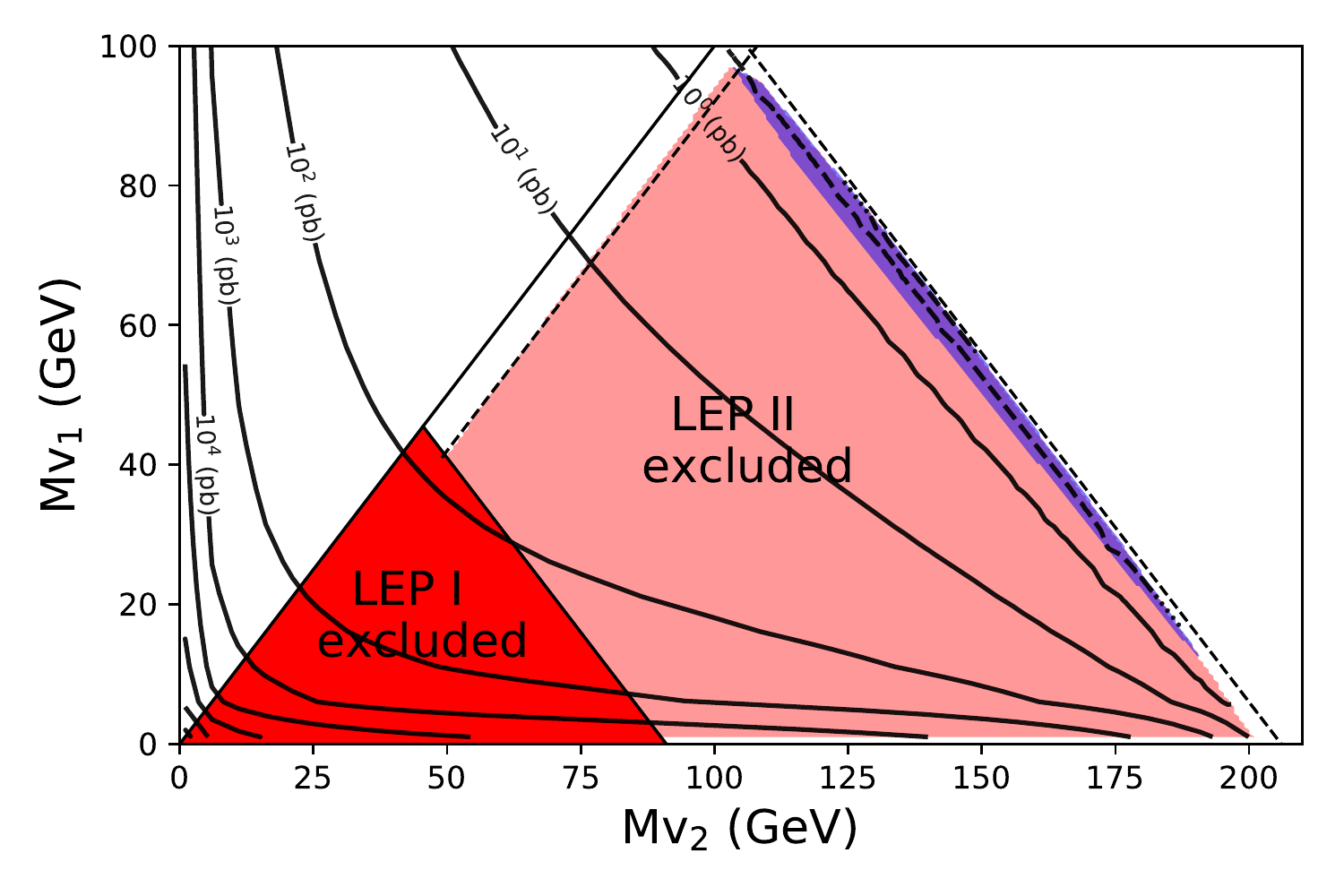}}%
{\includegraphics[width=0.50\textwidth]{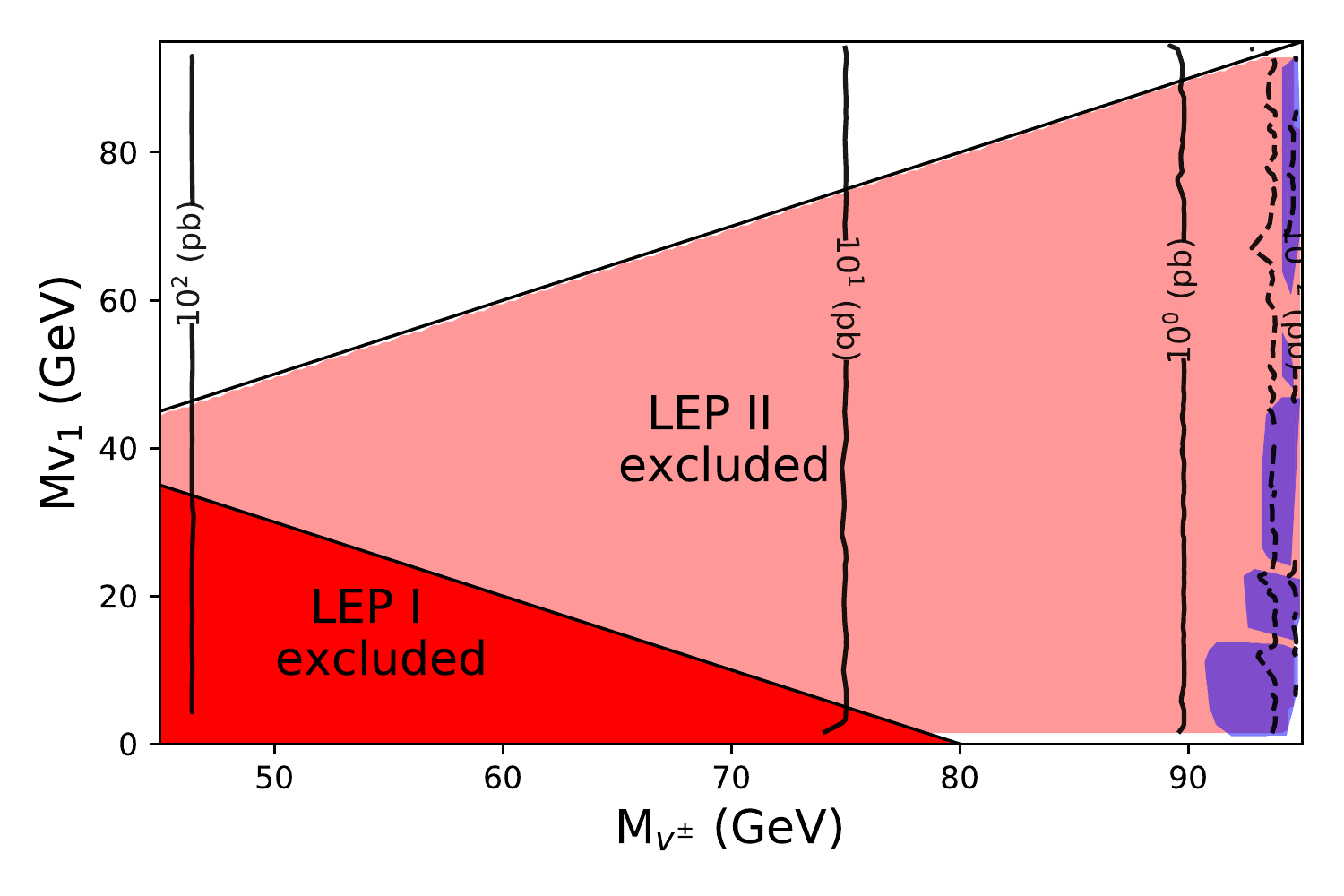}}%
\vskip -0.5cm\hspace*{-3cm}(a)\hspace*{0.48\textwidth}(b)
\caption{\footnotesize (a) Allowed mass region for neutral vectors based on 95\% C.L. upper-limits on $e^+e^-\rightarrow \tilde{\chi}_1^0 \tilde{\chi}_2^0$ cross section at $\sqrt{s} = 189$ GeV \cite{Abreu:2000dd}. The solid black contour lines indicate the production cross section $e^+e^-\rightarrow V^1V^2$ at LEP. The red(blue) zones are forbidden (allowed) by LEP-II data. The red shaded region is excluded by LEP-I data on the $Z$ boson width (see \ref{LEPI}). (b) Allowed mass region for charged and neutral vector based on 95\% C.L. upper-limits on $e^+e^-\rightarrow \tilde{\chi}_1^+ \tilde{\chi}_2^- $ cross section at $\sqrt{s} = 189$ GeV \cite{Abbiendi:1999ar}} \label{neu}
\end{figure}

In view of the identical topologies in the processes of the i2HDM and our model, it seems natural to extend the LEP bounds to our vectorial case. The concern is whether the efficiencies of the vectorial signals are similar to the SUSY ones. In the case of neutral state production, the process $e^+e^-\rightarrow Z\rightarrow V^1V^2$ shows a distribution more isotropic and similar to neutralinos, because both cases, having intrinsic spin, have the ability to conserve angular momentum. Scalars, on the other hand, which are produced through the same topology than the vector ones, $e^+e^-\rightarrow Z\rightarrow H^0A^0$, are produced in $p$-waves, making the scalars to have large transverse momentum. Additionally, as it has been shown in \cite{PhysRevD.79.035013}, the angular differences between SUSY signals and the scalar ones are even more reduced when are added the decay products of their respective new states. Therefore, we expect similar efficiencies among our signals and SUSY ones, allowing to recast LEP-II bounds.

In Fig. \ref{neu}(a) we show the recast limits from neutralinos searches at LEP-II to our model \cite{Abreu:2000dd}. The allowed region is a small narrow blue area close to the LEP energy threshold. The exclusion region is notoriously higher than the scalar case \cite{PhysRevD.79.035013} because the production cross section for vectors present an enhancement through their longitudinal polarization, compared to the scalar case (see \ref{fig:cs}). The resulting excluded region is
\begin{eqnarray}
 &\footnotesize \Mv[1] < 100 \text{ GeV}\quad \&\quad \Mv[2] < 200 \text{ GeV} \quad \&\quad \Delta M_{12}^- > 8 \text{ GeV} \quad \&\quad \Delta M_{12}^+ < E^{LEP} - 8 \text{GeV} \label{LEPIIa}
\end{eqnarray}
where $\Delta M_{12}^{\mp} \equiv \Mv[2] \mp \Mv[1]$ and $E^{LEP}$ is the maximum LEP center of mass energy ($189$ GeV).

Additionally, charginos searches \cite{Abbiendi:1999ar} also put strong constraints on the charged vector states $V^\pm$. As it is shown in Fig. \ref{neu}(b), the limit on the charged vectorial mass results to be
\begin{eqnarray}
 &\Mvc \lesssim 93 \text{ GeV}. \label{LEPIIb}
\end{eqnarray}
\subsection{$H\rightarrow \gamma \gamma$ constraints from LHC data}
%%%%%%%%%%%%%%%%%%%%%%%%%%%%%%%%%%%%%%%%%%%%%%%%%%%%%%%%%%%%%%%%%%%
In the SM, there is no interaction between the Higgs and photons at tree level, however, the Higgs boson can decay into a pair of photons due to one-loop processes which include the charged gauge bosons $W^{\pm}$ and fermions as internal particles. In this context our new doublet vector play an important role introducing new corrections to $\Gamma(H \rightarrow \gamma \gamma)$ through the charged vectors $V^{\pm}$ which can generate deviations to the diphoton rate predicted by the SM.
However, recent measurements (see, for instance, \cite{Aaboud:2018xdt}) show that the experimental
value of $\Gamma(H \rightarrow \gamma \gamma)$ is very close to the SM prediction, implying a  strong restrictions for models that go beyond the SM.

The partial width decay of the Higgs boson into two photons in the DVDM is
\begin{equation}
\Gamma(H \rightarrow \gamma \gamma) = \frac{\alpha_{em}M_H^3}{256\pi^3 v^2}\left|\sum_i N_{ci} Q_i^2 F_{1/2}(\beta_i) + F_1 (\beta_W) + \frac{\ld[2]}{2}\left(\frac{v}{\Mvc}\right)^2F_1(\beta_V)\right|^2,
\end{equation}
where $\alpha_{em}$ is the electromagnetic fine-structure constant, $M_H$ is the mass of the Higgs boson, $v$ is the 245 GeV Higgs field vev, $N_{ci}$, $Q_i$ and $\beta_i = 4M_i^2/M_H^2$ are the color factor, the electric charge and a dimensionless factor respectively for a certain $i$ fermion running in the loop. In the same way we define the dimensionless factors $\beta_W = 4M_W^2/M_H^2$ and $\beta_V = 4M_V^2/M_H^2$ for the charged $W^{\pm}$ boson and the new charged vector $V^{\pm}$ contributions in the loop respectively. The functions $F_i$ are loop factors for particles of spin given in the subscript:
\begin{equation}
F_{1/2} = -2\beta(1+(1-\beta)f(\beta)), \qquad F_1 = 2+3\beta + 3\beta(2-\beta)f(\beta), 
\end{equation}
with
\begin{equation*}
  f(\beta) = 
   \begin{cases}
     \arcsin(1/\beta)^2, & \text{for $\beta\geq$ 1}. \\
     -\frac{1}{4}\left[\ln{\frac{1+\sqrt{1-\beta}}{1-\sqrt{1-\beta}}}-i\pi \right]^2, & \text{for $\beta<$ 1.} 
   \end{cases}
\end{equation*}

We consider the most recent limit coming from the $\sqrt{s}$ = 13 TeV ATLAS Higgs data analysis \cite{Aaboud:2018xdt} to set restrictions on the parameter space. The new contributions respect to the SM are parametrized as the ratio of the branchig ratios between our model and the SM
\begin{equation}
\frac{Br^{BSM}(H\rightarrow \gamma\gamma)}{Br^{SM}(H\rightarrow \gamma\gamma)} = \mu^{\gamma\gamma} = 0.99 \pm 0.14. \label{mu_AA_val}
\end{equation}

The new contributions to $\mu^{\gamma\gamma}$ are governed by the parameters $\ld[2]$ and $\Mvc$ or, equivalently, by $\lamL$ and the difference of masses between $\Mv[1]$ and $\Mvc$, as previously shown in eq.(\ref{lambda-couplings}). 

\begin{figure}[htb]
\centering
{\includegraphics[width=0.50\textwidth]{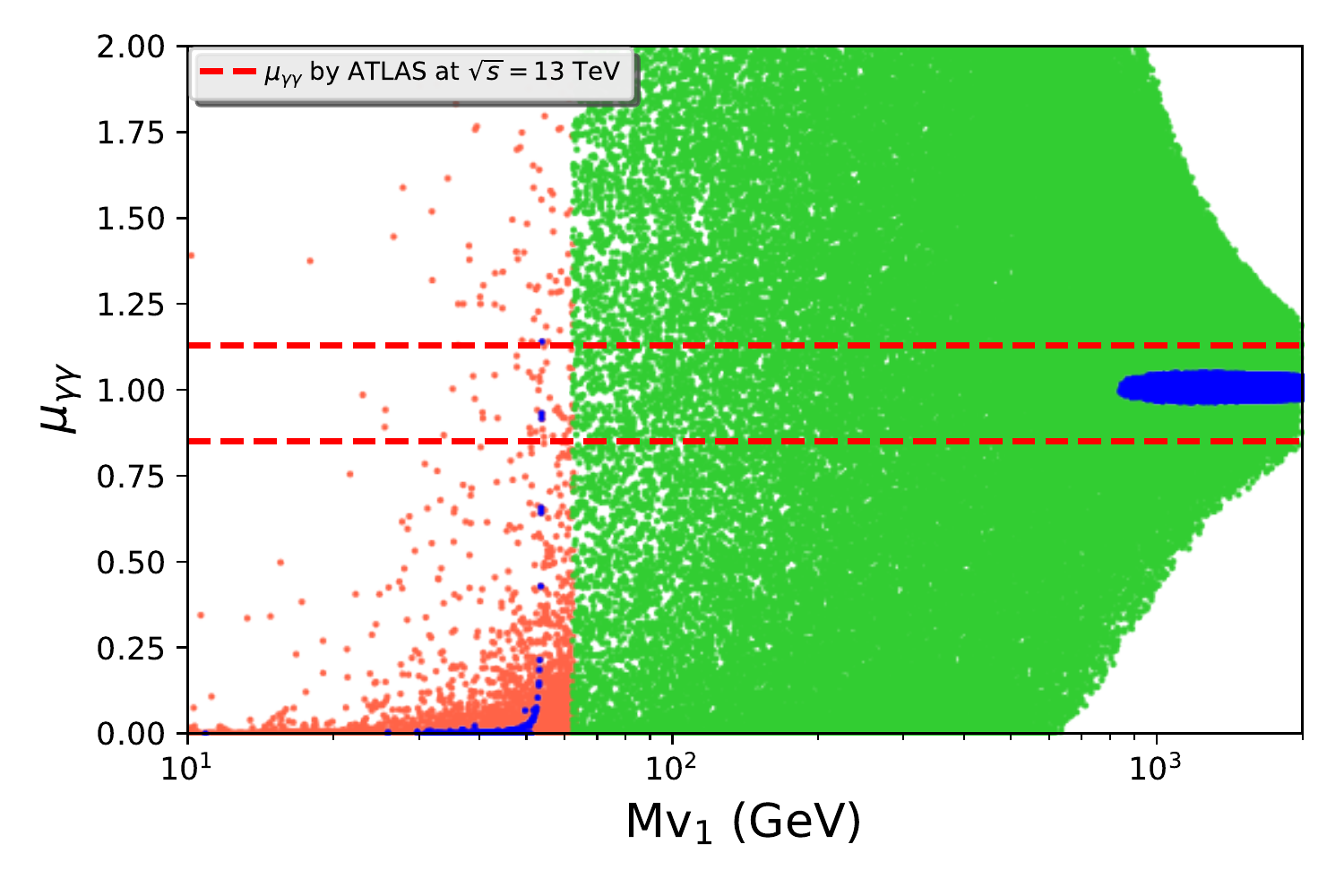}}%
{\includegraphics[width=0.50\textwidth]{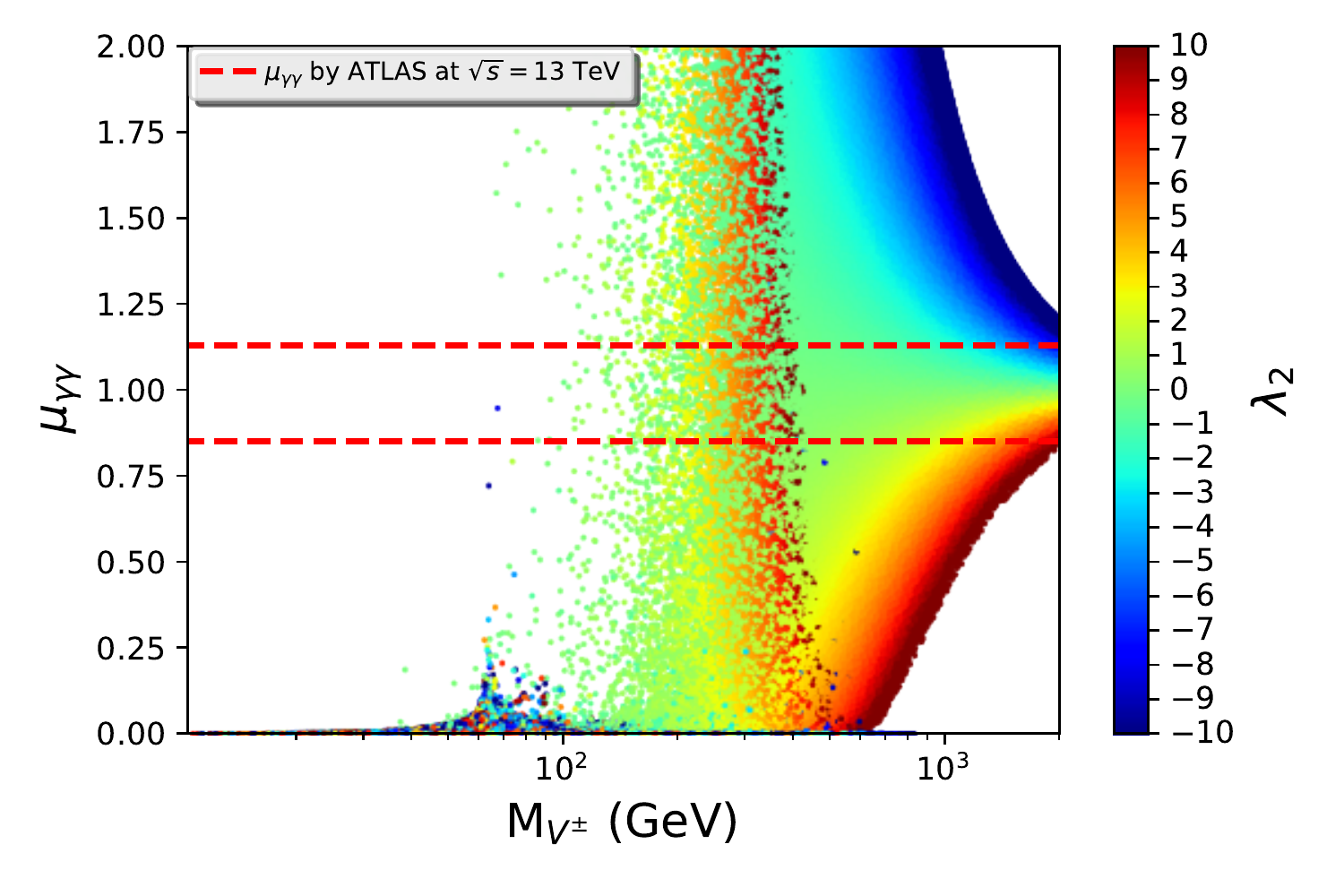}}%
\vskip -0.5cm\hspace*{-3cm}(a)\hspace*{0.48\textwidth}(b)
\caption{\footnotesize a) Diphoton rate $\mu^{\gamma\gamma}$ vs DM mass $\Mv[1]$ in two regions: the pink region correspond to $\Mv[1] \leq M_H/2$, where $H\rightarrow V^1V^1$ channel is open, and the green one to $\Mv[1]>M_H/2$, where $H\rightarrow V^1V^1$ channel is closed. The blue color represent relic saturation. b) Color map for diphoton rate as a function of the parameter $\ld[2]$ and the charged vector mass $\Mvc$. For these pictures we took into account perturbability restrictions (\ref{pert}). The horizontal red lines represent the global signal strength coming from $\sqrt{s}$ = 13 TeV ATLAS Higgs data analysis (\ref{mu_AA_val}). } \label{mu_AA}
\end{figure}

In Fig.(\ref{mu_AA})(a) we present the diphoton rate as a function of the DM mass $\Mv[1]$ where the parameter space was divided in two regions: the pink points $(10\leq \Mv[1] \leq M_H/2)$ represent the zone where the decay mode $H\rightarrow V^1V^1$ is open making the decay mode $H\rightarrow \gamma\gamma$ very low and therefore pushing the $\mu^{\gamma\gamma}$ under the experimental limit for most of the points, and the green points $(\Mv[1]>M_H/2)$ represent the zone where the decay mode $H\rightarrow V^1V^1$ is closed. In both regions we show in blue the points which are consistent with the observed amount of DM. We also present a color map of the parameter $\ld[2]$ as a function of the diphoton rate vs charged vector mass $\Mvc$ in Fig.(\ref{mu_AA})(b). In both cases the horizontal red lines represent the global signal strength coming from $\sqrt{s}$ = 13 TeV ATLAS Higgs data analysis (\ref{mu_AA_val}).

We can notice that diphoton rate constraints are very restrictives ruling out an important amount of the parameter space mostly when $|\ld[2]|$ takes big values in the region $\Mvc \gtrsim 400$ GeV. However, for higher masses such as $\Mvc \gtrsim 1$ TeV, still there is a region where $\mu^{\gamma\gamma}$ is within the experimental limit for high couplings, e.g. $|\ld[2]|>5$. Another interesting feature of the model is that the low mass region that can satisfied the PLANCK limit for $\RD$ is practically ruled out. On the other hand the high mass region which saturates the PLANCK limit  matches perfectly with the $\mu^{\gamma\gamma}$ measurements where $(|\ld[2]|<2)$ and ($\Mvc-\Mv[1] \lesssim 20$ GeV) values are preferred.

%%%%%%%%%%%%%%%%%%%%%%%%%%%%%%%%%%%%%%%%%%%%%%%%
\subsection{Invisible Higgs decay from LHC data}
%%%%%%%%%%%%%%%%%%%%%%%%%%%%%%%%%%%%%%%%%%%%%%%%
The Higgs boson is one of the portals connecting the dark sector with the SM, however there is an important restriction that we need to worry about. When $\Mv[1]\leq M_H/2$, the SM-Higgs boson can decay into Dark Matter particles, which translate into invisible decays. On the other hand, both ATLAS and CMS experiments at the LHC has been seaching for Higgs invisible decays at $\sqrt{s}= 7$, $8$ and $13$ TeV, putting the restrictive upper limit
\begin{eqnarray} \label{eq:cms}
Br(H \rightarrow \text{inv}) < 24\% ,  \label{inv_higgs}
\end{eqnarray}
at a 95\% of confidence level \cite{Aad:2015txa}\cite{Khachatryan:2016whc}. In this section we interpret the CMS upper bound as the maximum possible branching ratio of the Higgs boson into dark matter particles, i.e.
\begin{eqnarray}\label{eq:br}
 Br(H\rightarrow V^1V^1) \equiv \frac{\Gamma(H\rightarrow V^1V^1)}{\Gamma_{SM} + \Gamma(H\rightarrow V^1V^1)} < Br^{max}_{inv},
\end{eqnarray}
where $Br^{max}_{inv} = 24\%$, and $\Gamma_{SM}$ corresponds to the the full decay width of the SM Higgs. In our model, the decay width of the Higgs to two dark matter particles is given by
\begin{equation} \label{eq:pw}
\Gamma(H\rightarrow V_1V_1) = \frac{M_W^2\lamL^2}{8\pi g^2 M_H}\left(3 - \frac{M_H^2}{\Mv[1]^2} + \frac{1}{4}\frac{M_H^4}{\Mv[1]^4}\right)\sqrt{1-\frac{4\Mv[1]^2}{M_H^2}},
\end{equation}
where $g$ is the weak coupling constant. Replacing \ref{eq:pw} into \ref{eq:br} and solving for $\lambda_L$, we found the following constraint
\begin{eqnarray} \label{lamres}
|\lamL| < \left(\frac{8\pi \Gamma_{SM} g^2 M_H\left(\frac{1}{Br^{\text{max}}_{\text{inv}}}-1\right)^{-1}}{M_W^2\left(3 - \frac{M_H^2}{\Mv[1]^2} + \frac{1}{4}\frac{M_H^4}{\Mv[1]^4}\right)\sqrt{1-4\Mv[1]^2/M_H^2}}\right)^{1/2}.
\end{eqnarray}
This bound is extremely restrictive because it allows only for very small values of $\lamL$ \footnote{This strong contraint in the coupling among the Higgs boson and the dark matter is also shown in the i2HDM \cite{Belyaev:2018ext} with similar results.}. For example,  when $\Mv[1]$ is close to $M_H/2$ ($\sim$ 60 GeV), relation (\ref{lamres}) sets $\lamL \lesssim$ 0.03. This constraints is complementary to the one given by Higgs diphoton decay, which strongly constrained dark matter masses below $M_H/2$, eliminating almost completely the region $\Mv[1]\leq M_H/2$.

The case described above was based on the assumption that the sole channel contributing to the Higgs invisible decay is $H\rightarrow V^1V^1$. However, when $M_{V^2}< M_H/2$,  the channel $H\rightarrow V^2V^2$ can also contributes to the invisible Higgs decay provided that $\Delta M = M_{V^2}-M_{V^1}$ is small enough (of the order of a few GeV or less), to forbid $V^2$ to decay into $V^1$ and a detectable pair of fermions. Considering that $\lambda_R = \lambda_L + \frac{2\left(M_{V^2}^2 - M_{V^1}^2\right)}{v^2}$, and in this case, $M_{V^2} \approx M_{V^1}$, then $\lambda_L \approx \lambda_R$. Therefore, in this case the limit on $\lambda_L$ can be easily modified.

Finally, in the case of a small $V^\pm -V^1$ mass split, the channel $H\rightarrow V^\pm V^\mp$ may also contributes to the Higgs invisible decay channel. However, LEP limits \ref{LEPII} put very strong constraints on the allowed masses of the charged vectors, then making the Higgs decay into the on-shell charged vectors kinematically forbidden. 

%%%%%%%%%%%%%%%%%%%%%%%%%%%%%%%%%%%%%%
\subsection{Relic Density constraints}
%%%%%%%%%%%%%%%%%%%%%%%%%%%%%%%%%%%%%%
As we mentioned in section \ref{DMM}, our model has a 6-dimensional parameter space but only four free parameters are relevant for our study: three physical masses of the vector fields ($\Mv[1], \Mv[2], \Mvc$),and one coupling constant ($\lamL$) between the SM-Higgs boson and $V^1$. In order to show a general qualitative description of the dark matter relic density $\RD$ as a function of the parameter space we can fix some of them and perform a scan over the more relevant ones. The result should be in agreement with the WMAP \cite{Hinshaw:2012aka} and PLANCK \cite{Ade:2013zuv, Planck:2015xua} measurements:
\begin{equation}
\RD = 0.1184 \pm 0.0012.  \label{PLANCK}
\end{equation}
The interaction between both the dark sector and the SM is through the SM-Higgs boson and the electroweak gauge bosons, however the interaction with the latter it is fixed by gauge couplings. For simplicity we will consider as well $\Mv[2]=\Mvc$ \footnote{This equivalently to do $\ld[3] = \ld[4]$ as you can easily check from\ref{MassV2}.}. Therefore the two relevant parameters are ($\Mv[1]$, $\lamL$).  

\begin{figure}[htb]
\centering
{\includegraphics[width=0.5\textwidth]{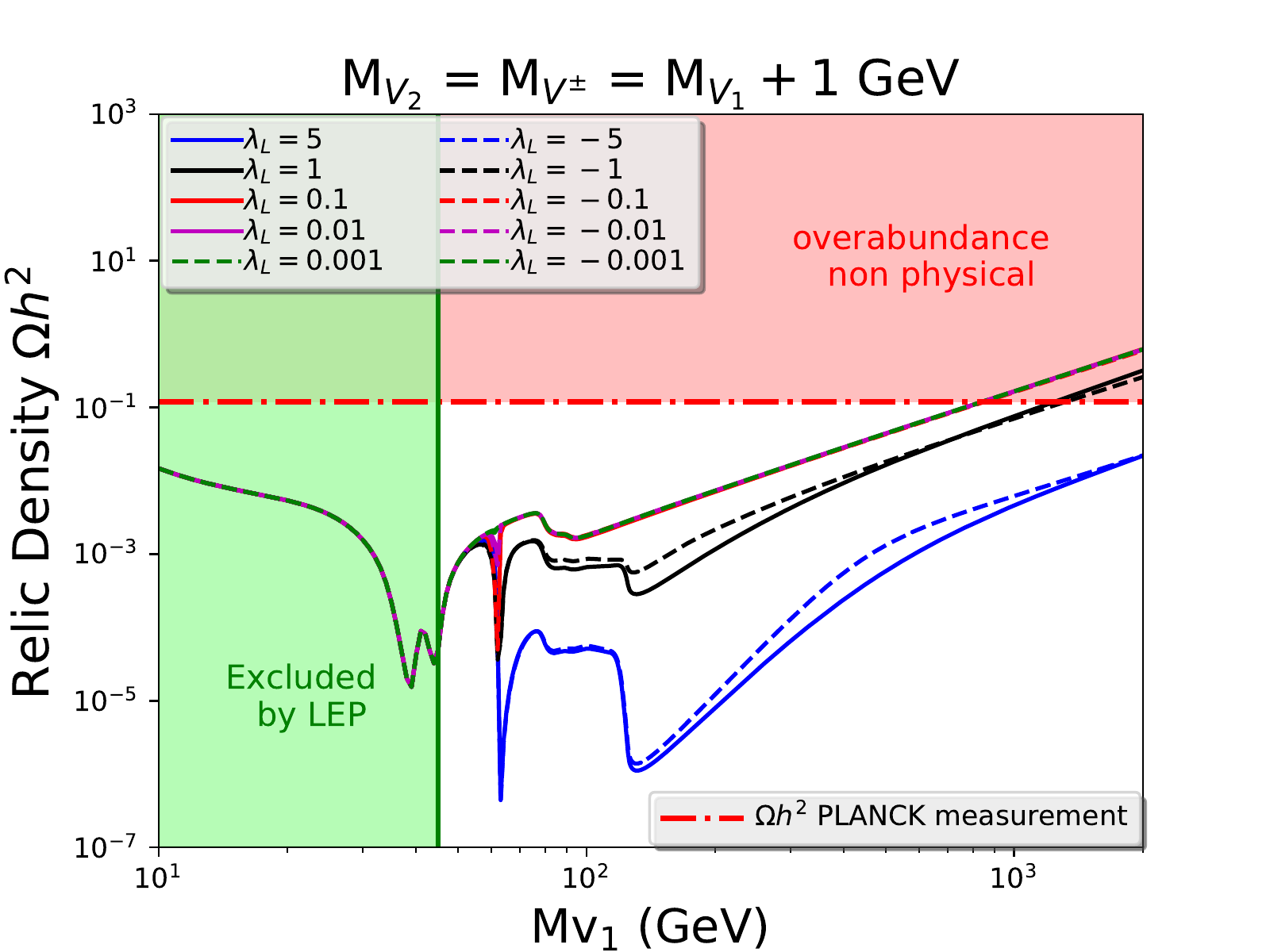}}%
{\includegraphics[width=0.5\textwidth]{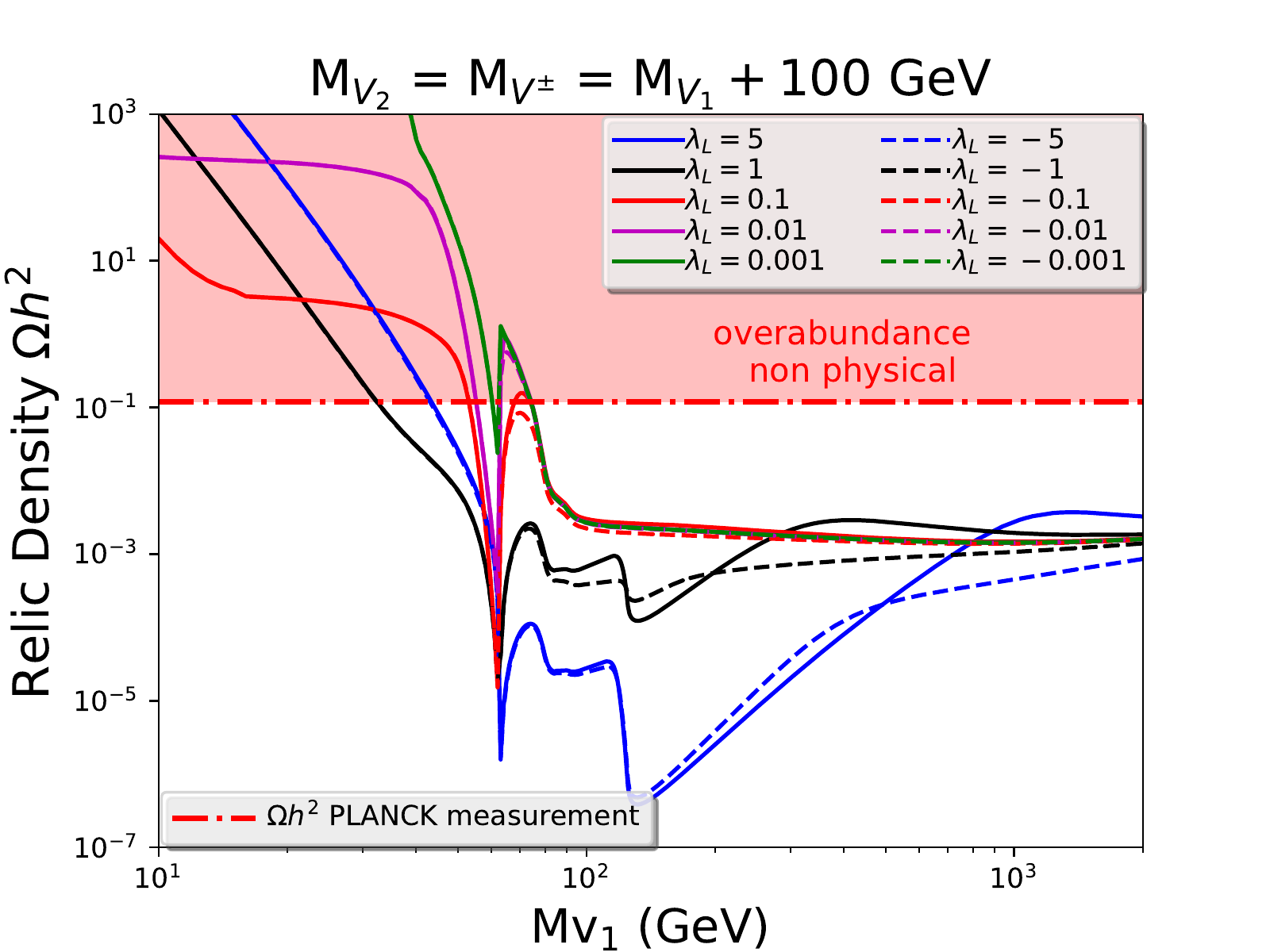}}%
\vskip -0.5cm\hspace*{-3cm}(a)\hspace*{0.48\textwidth}(b)
\caption{\footnotesize Relic density $\RD$, as a function of $\Mv[1]$ for different values of $\lamL$ in a quasi-degenerate scenario (a) where $\Mv[2]=\Mvc = \Mv[1] + 1$ GeV and a no-degenerate scenario (b) where $\Mv[2]=\Mvc = \Mv[1] + 100$ GeV. The horizontal red line corresponds to the central value of the relic density measured by PLANCK. The green area indicate the excluded region by LEP measurements.} \label{fig:RD-mv1-lamL}
\end{figure}

We will present two characteristic scenarios which we will refer to as: a) \textit{quasi-degenerate} case, where $\Delta M = \Mv[2]-\Mv[1] = 1$ GeV, and b) the \textit{non-degenerate} one, in which $\Delta M= \Mv[2]-\Mv[1] = 100$ GeV. In Figure \ref{fig:RD-mv1-lamL} we present a 2-dimensional parameter space where we show the $\RD$ as a function of the DM mass $\Mv[1]$ for different values of $\lamL$ in the two scenarios mentioned above. The horizontal red dashed line corresponds to the central value of the relic density measured by PLANCK.

The first important aspect we can appreciate of this model is that there are two regions in which it can fulfill the DM budget. The first saturation zone happens between $30 < \Mv[1] < 80$ GeV for a non-degenerate scenario, as we can see from Fig.(\ref{fig:RD-mv1-lamL})(b). In this case the main mechanism of annihilation is through s-channel Higgs boson exchange which is controlled by the $\lamL$ coupling. Interestingly, there is a considerable area of overabundance for small values of $\Mv[1]$ even for large values of $\lamL$. Of course, this region must be excluded as non physical.

 The second saturation region takes place when $\Mv[1]>830$ GeV in the quasi-degenerate scenario (see Fig.(\ref{fig:RD-mv1-lamL})(a)). In this zone the interaction between the DM and the longitudinal polarization of $W^{\pm}$ and $Z$ boson becomes dominant. This interaction is modulated by $\lambda_i$ quartic couplings which in turn depend on the mass difference among the new vectors as it is shown in eq.(\ref{lambda-couplings}). When $\Delta M$ is small the $\lambda_i$ become small enough to produce a suppression in the annihilation average cross section for these channels pushing the DM abundance up to reach the saturation limit, even when the (co)annihilation effects are present which become subdominant. In contrast, in the non-degenerate cases the annihilation of DM is more efficient due to the large values of $\lambda_i$ which results in the asymptotically flat behavior of abundance for high DM mass values.

The overabundance seen in the non-degenerate scenario for small values of $\Mv[1]$ completely disappears in the quasi-degenerate case due to effects of (co)annihilation which introduces new sources of annihilation of DM, pushing the abundance below the PLANCK experimental limit. When $\Mv[1] \sim 40$ and $\Mv[1] \sim 45$ GeV we can note the effects of resonant (co)annihilation through $V^1 V^{\pm}\rightarrow W^{\pm}$ and $V^1V^2\rightarrow Z$ channels respectively that manifest on the Fig.(\ref{fig:RD-mv1-lamL})(a) as two inverted peaks. 

At exactly $\Mv[1] \sim 62.5$ GeV the resonant annihilation through the Higgs boson take place as we can see in both scenarios as a deep peak. After that resonance we observe three points where the abundance of DM decreases considerably. This happens markedly at $\Mv[1] \sim 80$ GeV through the opening channel $V^1 V^1 \rightarrow W^+W^-$ and more tenuously at $\Mv[1] \sim 90$ GeV through $V^1 V^1 \rightarrow ZZ$. Finally at $\Mv[1] \sim 125$ GeV the opening of $V^1 V^1 \rightarrow HH$ take place corresponding to the reduction of DM relic density through s-channel Higgs boson. 

One can also observe that in the case of $\Delta M = 100$ GeV, for $M_{V^1}$ below 65 GeV, DM co-annihilation is suppressed and the relic density is equal or below the experimental limit only for large values of $\lambda_L$ ($\lambda_L > 0.1$) which are excluded by LHC limits on the invisible Higgs decay.

Finally it is easy to notice that for larger values of $\lamL$ the abundance of DM decreases, however, is important to stress that there is a slight difference for the case in which $\lamL$ takes positives and negatives values after $\Mv[1]\sim 62.5$ GeV. This behavior is due interference effect between the s-channel Higgs boson exchange diagram and and those involving gauge bosons.

%%%%%%%%%%%%%%%%%%%%%%%%%%%%%%%%%%%%
\subsection{Direct Detection limits}
\begin{figure}[htb]
\centering
{\includegraphics[width=0.5\textwidth]{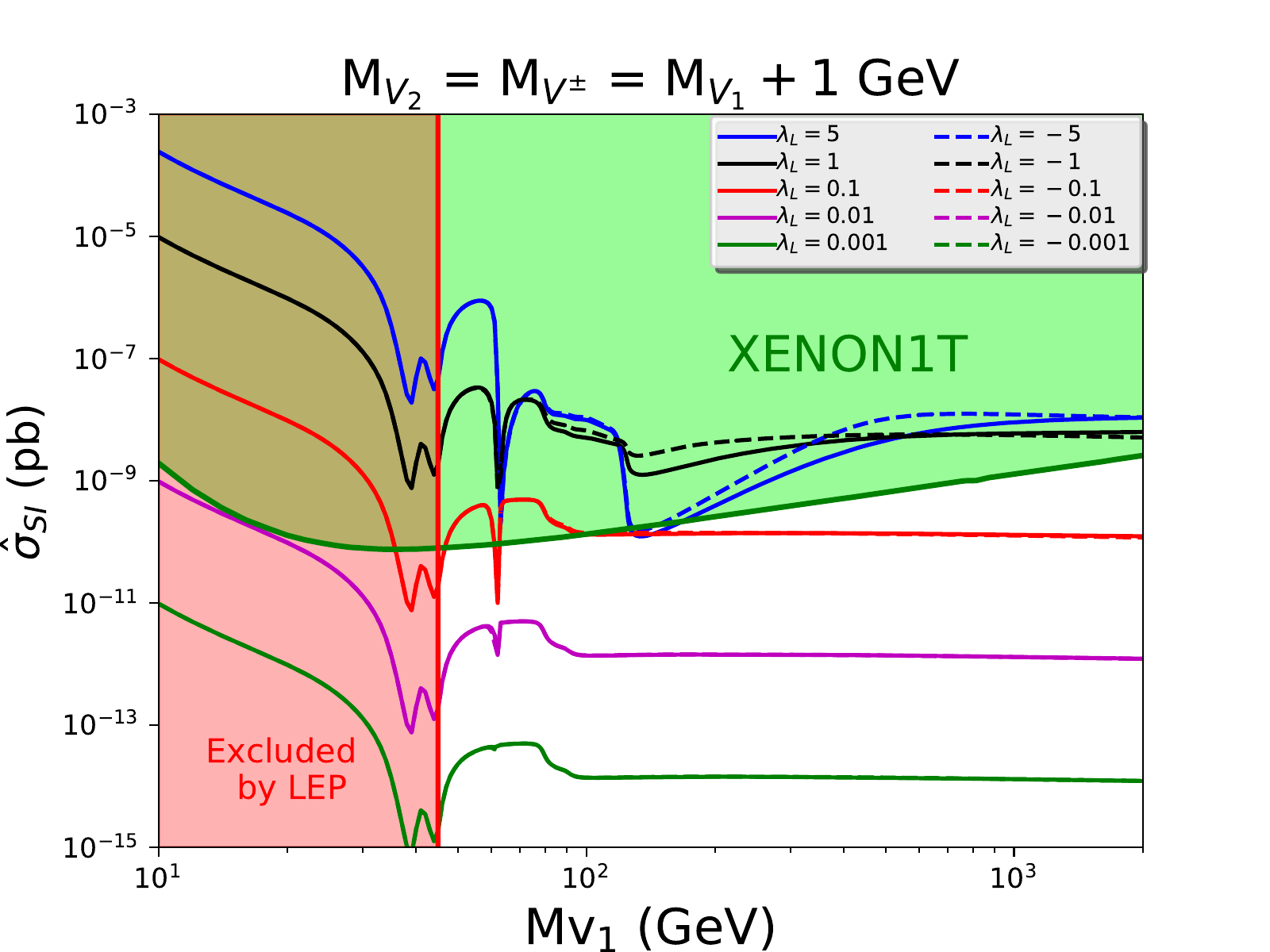}}%
{\includegraphics[width=0.5\textwidth]{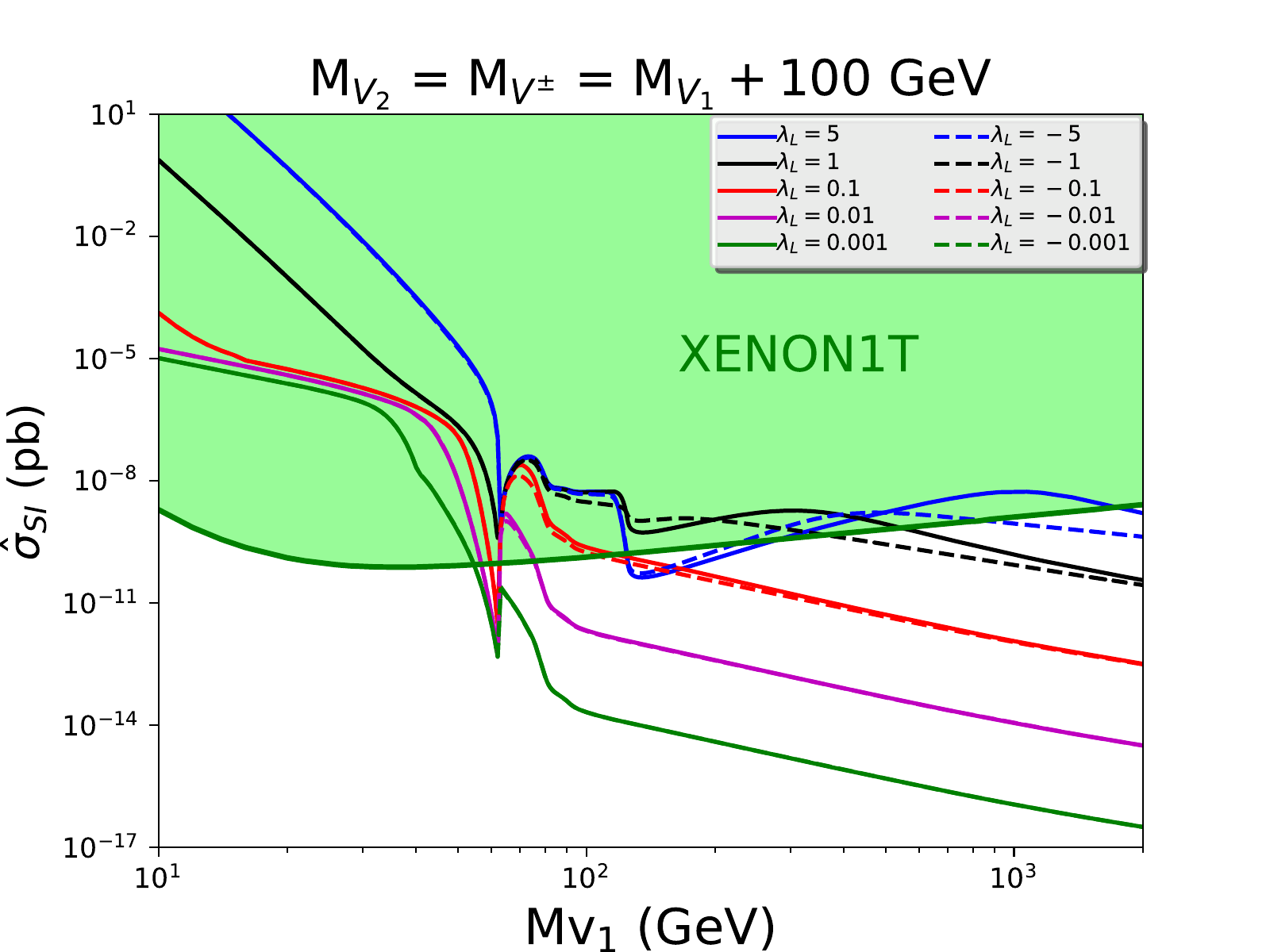}}%
\vskip -0.5cm\hspace*{-3cm}(a)\hspace*{0.48\textwidth}(b)
\caption{\footnotesize Rescaled spin independent direct detection cross section $\hat{\sigma}_{\text{SI}}$ versus $\Mv[1]$ and the XENONT1 constraint for several values of $\lamL$. The red-shaded region in the left frame is excluded by LEP data.} \label{fig:DD-mv1-lamL}
\end{figure}

%%%%%%%%%%%%%%%%%%%%%%%%%%%%%%%%%%%%
We consider as well whether our model is consistent with limits coming from XENON1T \cite{Aprile:2017iyp} experiment studying the rescaled spin independent proton-DM scattering cross section 
\begin{equation}
\hat{\sigma}_{SI} = (\Omega_{DM}/\Omega_{\text{PLANCK}})\times \sigma_{SI}(V^1 p\rightarrow V^1 p)\end{equation}
which allows us to take into account the case when the vector $V_1$ contribute only partially to the total amount of DM. This approach is useful to take into account other sources that can contribute to fulfill the DM budget. We present the $\hat{\sigma}_{SI}$ as a function of the DM mass for several values of $\lamL$ in the quasi-degenerate and non degenerated scenario as we shown in Figure\ref{fig:DD-mv1-lamL}. The green area, shown in both plots is the excluded region from the direct detection (DD) experiment and the soft red color in Figure\ref{fig:DD-mv1-lamL}(a) is excluded by LEP data.

The $\sigma_{SI}$ is through the t-channel with the Higgs boson as a mediator, therefore we can notice immediately that $\lamL$ plays an important roll which is scale the strength of the interaction between DM and nucleus of ordinary matter. In the quasi-degenerated scenario the asymptotically flat behavior of the $\hat{\sigma}_{SI}$ for $\Mv[1]> 100$ GeV can be explained because as $\Mv[1]$ take higher values, the cross section $\sigma_{SI}$ is decreasing, however this effect is compensated by the fact that there is more abundance of DM as the value of $\Mv[1]$ is increasing. We can check this from Fig\ref{fig:RD-mv1-lamL}(a). On the other hand, in the non-degenerate scenario the $\RD$ is relatively constant after the DM annihilation channel $V^1V^1\rightarrow HH$ is opened (Fig\ref{fig:RD-mv1-lamL}(b)), therefore, as the value of DM mass is increasing the $\hat{\sigma}_{SI}$ is taking smaller values.

%%%%%%%%%%%%%%%%%%%%%%%%%%%%%%%%%%%%%%%%%%%%%%%%%%%%
\section{Dark matter phenomenology} \label{DM_pheno}
%%%%%%%%%%%%%%%%%%%%%%%%%%%%%%%%%%%%%%%%%%%%%%%%%%%%
The previous description provides us with a qualitative overview of the parameter space. However, in order to have a deeper understanding of the model we perform a random scan using 7 million points of the most relevant parameters that have direct interference in the phenomenology of dark matter. The range of the parameters used in the scan can be summarized in table(\ref{tab_param}).

\begin{table}[ht]
	\caption{\footnotesize Range of the 4-dimensional parameter space.}
	\vspace{-0.5cm}
	\begin{center}
		\begin{tabular}{c|c|c}
			\hline
			\hline
			\textbf{Parameter}  & \textbf{min value} &  \textbf{max value} \\ \hline \hline 
			$\Mv[1]$ [GeV]   &  10  &   2000  \\ %\hline
			$\Mv[2]$ [GeV]   &  10  &   2000  \\ %\hline
			$\Mvc$ [GeV]     &  10  &   2000  \\ %\hline
			$\lamL$          &  -12 &   12  \\ \hline
		\end{tabular} 
	\end{center}
	\label{tab_param}
\end{table}

\begin{figure}[htb]
\centering
{\includegraphics[width=0.51\textwidth]{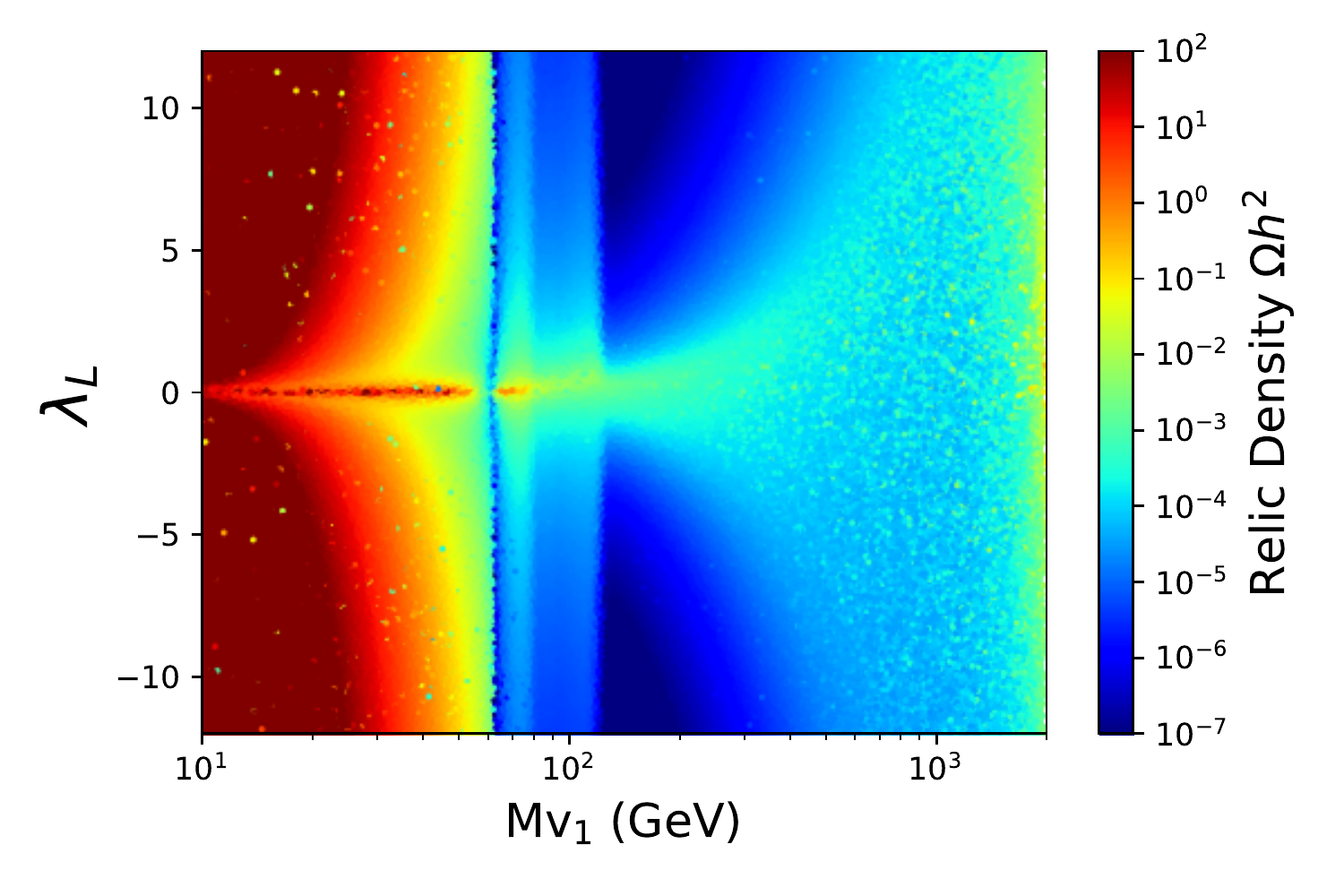}}%
{\includegraphics[width=0.51\textwidth]{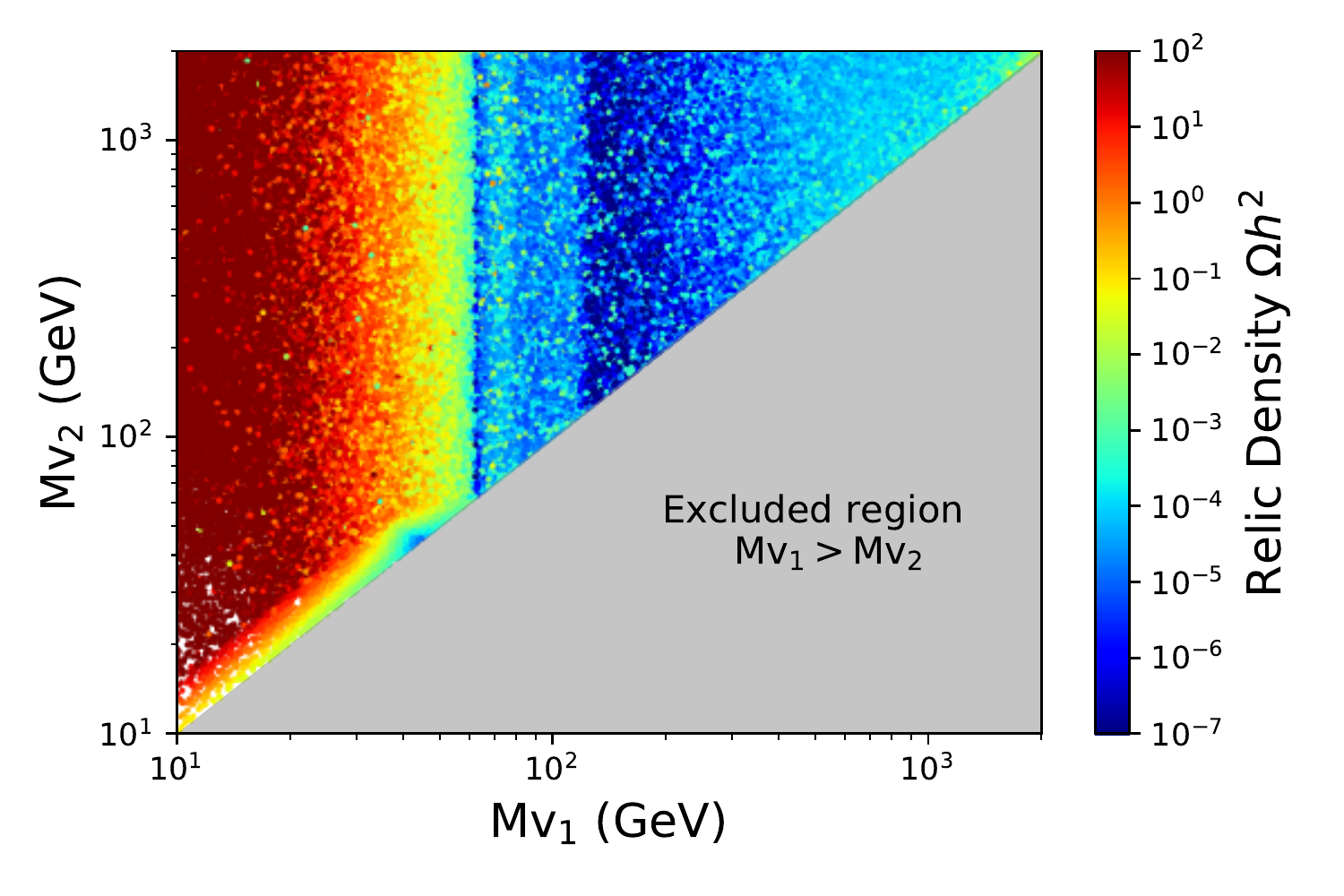}}%
\vskip -0.5cm\hspace*{-3cm}(a)\hspace*{0.48\textwidth}(b)
\centering
{\includegraphics[width=0.51\textwidth]{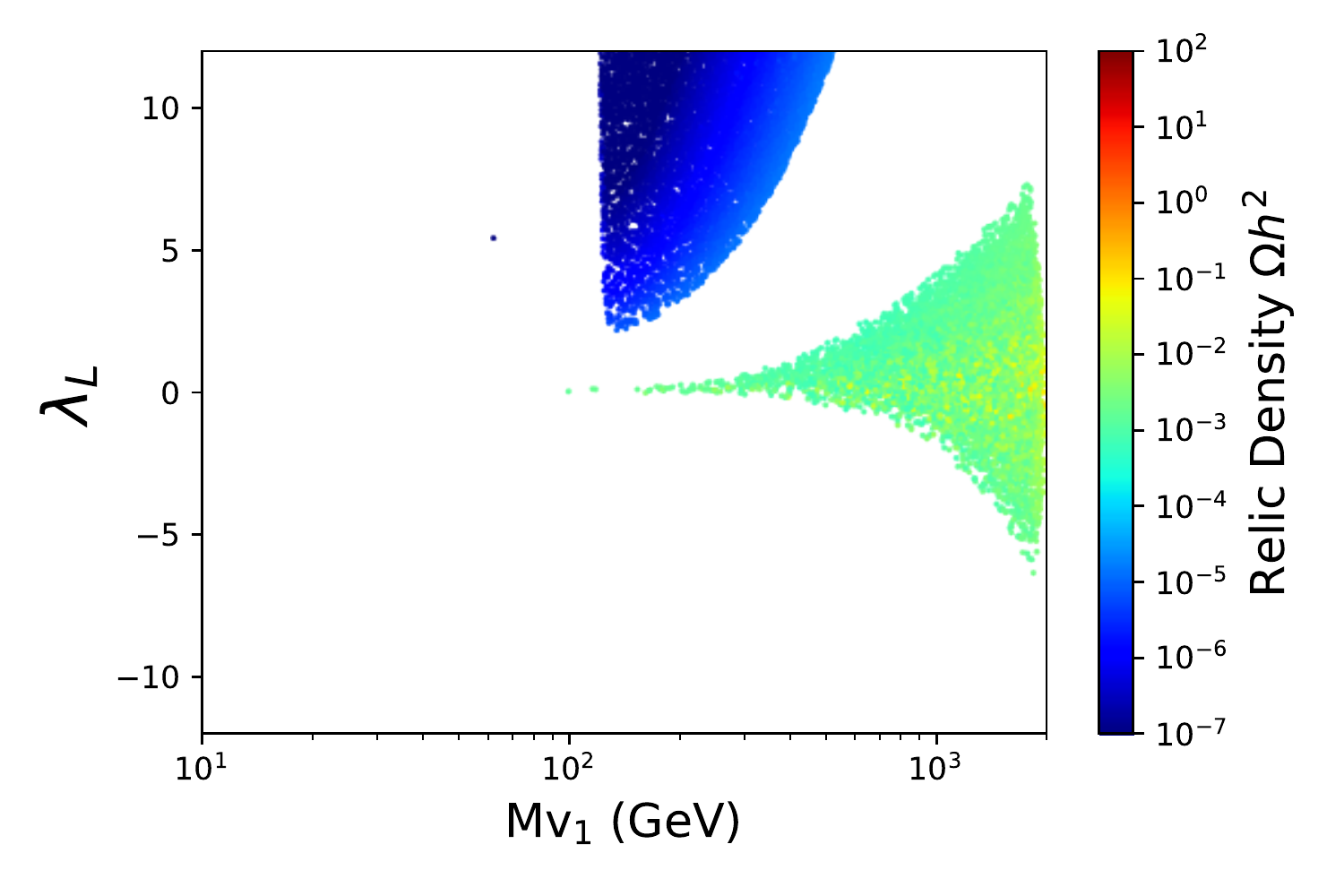}}%
{\includegraphics[width=0.51\textwidth]{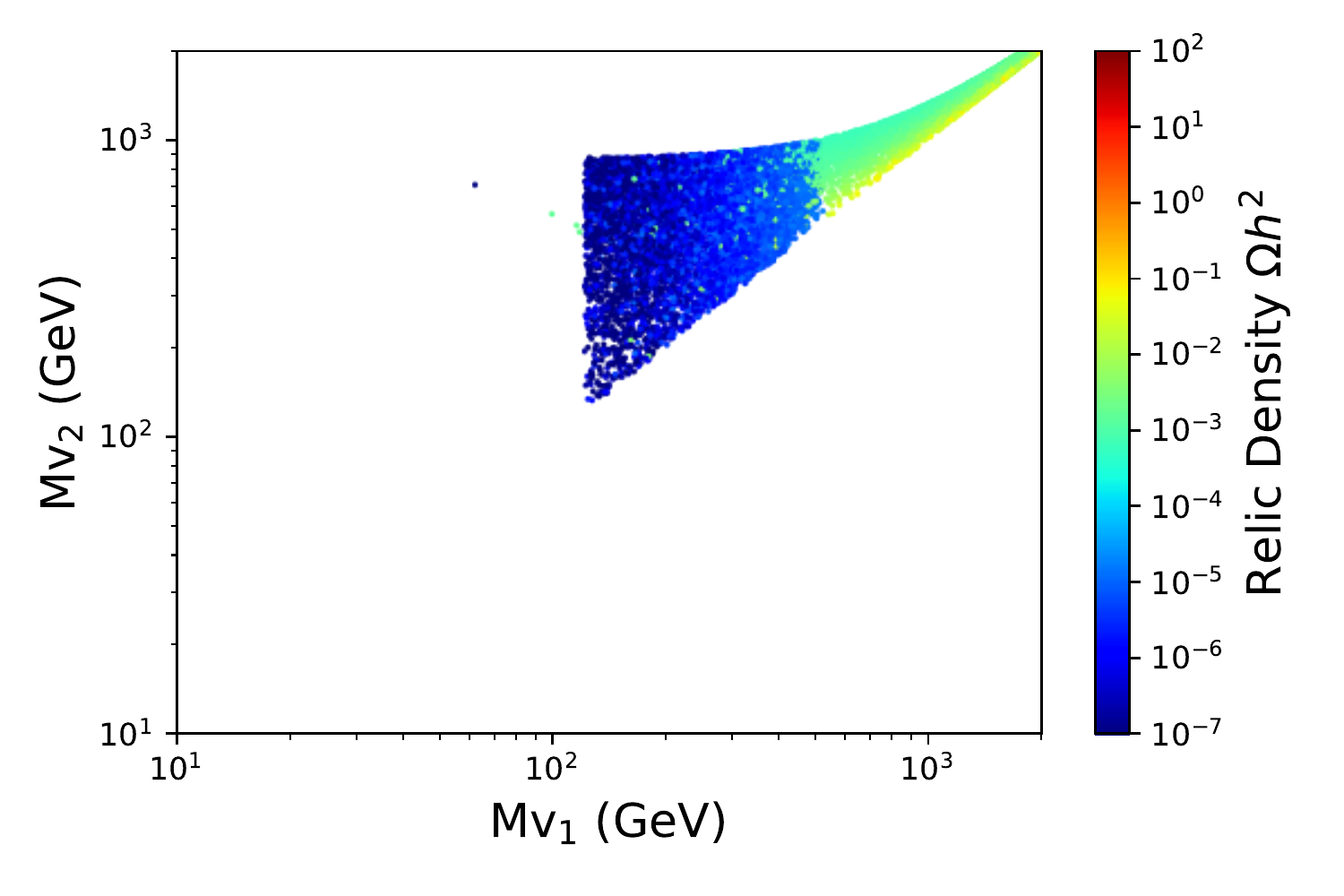}}%
\vskip -0.5cm\hspace*{-3cm}(c)\hspace*{0.48\textwidth}(d)
\caption{\footnotesize 2-D projections of the 4-dimensional parameter space presented as a color map of $\RD$ in two different planes: ($\Mv[1],\lamL$) plane for Fig.(a,c) and ($\Mv[1],\Mv[2]$) plane for Fig.(b,d). In the first row we present the parameter space without any constraint and in the second one we applied all the theoretical and experimental constraints with exception of  DM under-abundance.} \label{fig:full-scan}
\end{figure}

The result of our scan is presented in Fig. (\ref{fig:full-scan}) where we show several plots with 2-D projections of the 4-dimensional parameter space as a color map of DM Relic Density.
% for the planes ($\Mv[1],\lamL$) in Fig.(\ref{fig:full-scan})(a,c) and ($\Mv[1],\Mv[2]$) for Fig.(\ref{fig:full-scan})(b,d). 
We considered the parameter space without any theoretical or experimental constraint in the first row, and then, in the second row we took into account Perturbativity (\ref{pert}), LEP limits (\ref{LEPI},\ref{LEPIIa} and \ref{LEPIIb}), Higgs decay into two photons (\ref{mu_AA_val}), Invisible Higgs decay (\ref{inv_higgs}), overabundance DM Relic density (\ref{PLANCK}) and Xenon1T Direct Detection constraints.

As we explained previously, and without losing generality,
 %there is not an observable that can distinguish between $V^1$ and $V^2$. 
 we work in the region where $\Mv[1]<\Mv[2]$ and therefore $\ld[4]>0$. 
 % If we consider $\ld[4]<0$ this choice is not a new region since it is equivalent to the positive $\ld[4]$ situation. We can perform a transformation of the doublet by a phase shift $iV_{\mu}$ and swap $V^2 \leftrightarrow V^1$ without any observable effect. 
 For this reason we exclude the region $\Mv[1]>\Mv[2]$ as we can see from the gray region in Fig. (\ref{fig:full-scan})(b).

The different pattern of colors represent the amount of DM that the model is capable of explain considering a thermal production mechanism, where the dark red color in the low DM mass region ($\Mv[1]\lesssim 45$ GeV) of Fig.(\ref{fig:full-scan})(a,b) represent over-abundance which we consider as non physical. The dark blue color are the regions with extreme under-abundance of DM which is more accentuated for large values of $\lamL$ in the zone where $\Mv[1]>M_H/2$ after the respective annihilation channels ($WW$, $ZZ$ and $HH$) are progressively opened, reflecting the same pattern shown previously in figure(\ref{fig:RD-mv1-lamL}).

Looking at Figure (\ref{fig:full-scan})(a,b), the resonant annihilation through the Higgs boson is easily recognized by the vertical separation around $\Mv[1]\sim 62.5$ GeV where a steep break in the color pattern can be seen, changing from an light green to a blue Dark. We can also notice the resonant (co)annihilation through the $Z$ boson in the plane ($\Mv[1],\Mv[2]$) of Fig.(\ref{fig:full-scan})(b) at the region $\Mv[1] = \Mv[2] \sim 45$ GeV. 

Taking into account perturbative restrictions, the region of the parameter space that shows an important mass difference between $\Mv[1]$ and $\Mv[2]$ is excluded since this large difference increases the values of the quartic coupling beyond the allowed value set by (\ref{pert}). This effect can be seen clearly in Fig. (\ref{fig:full-scan})(d) where the region with $\Mv[2]>900$ GeV for $\Mv[1]< 500$ GeV is excluded. Only when the mass difference becomes relatively small, $\Mv[2]$ can admit larger values.

By incorporating the restrictions coming from Higgs invisible decay almost all the parameter space for $\Mv[1]\lesssim M_H/2$ disappears with exception of a very narrow region where $\lamL$ parameter take small values ($\lamL\lesssim 0.02$). This happen because the dominant annihilation channel is through the higgs boson exchange.

The Higgs diphoton rate (\ref{mu_AA_val}) introduce strong restrictions on the parameter space specially for negative values of $\lamL$. We can see that restriction in the Fig.(\ref{fig:full-scan})(c) where $\lamL$ is limited from below through the parabolic shape as we increase the values of $\Mv[1]$. The diphoton rate depend explicitly on $\ld[2] = \lamL +2(\Mv[1]^2-\Mvc^2)/v^2$, where the difference of squared masses is always negative because $\Mv[1]<\Mvc$, therefore when the mass difference is large and $\lamL$ takes high negative values, the parameter $\ld[2]$ grows in demacy, causing a great deviation from the experimental value of $\mu^{\gamma\gamma}$, this can also be seen as well in Fig. (\ref{mu_AA})(b).

The additional constraint from XENON1T DD experiments removes part of the parameter space contained between $63<\Mv[1]<125$ GeV where the direct detection rate is more sensitive. This  affect the region for positive and negative values of $\lamL$, however the negative part was removed previously by the Higgs diphoton rate constraint as we can see from Fig.(\ref{fig:full-scan})(c). The scattering cross section between $V_1$ and nuclei is through the t-channel with the Higgs boson as a mediator, therefore it depends explicitly on the parameter $\lamL$. For large values of $\lamL$ the abundance of DM is low, but not low enough to suppress the DM detection rate through DD signal. Only when $\lamL$ is small ($\lesssim 0.02$), the region between $90<\Mv[1]<200$ GeV of the parameter space is able to bypass the limits of direct detection. When we move to a high DM mass region ($\Mv[1]\gtrsim 200$ GeV), where the DD rate is less sensitive, we still have a excluded region with parabolic shape that it is only reached for large values of $\lamL$. It produces a clear division between a low density of DM zone with the rest of the parameter space. However, in the case of high degeneracy among the vector masses for the region $\Mv[1]>900$ GeV the DD rate is able to restrict parameter space for values of $\lamL$ up to 1, as we will see later in the next subsection. 

\subsection{Vector Dark Matter as the only source}
%%%%%%%%%%%%%%%%%%%%%%%%%%%%%%%%%%%%%%%%%%%%%%%%%%

\begin{figure}[htb]
\centering
{\includegraphics[width=0.50\textwidth]{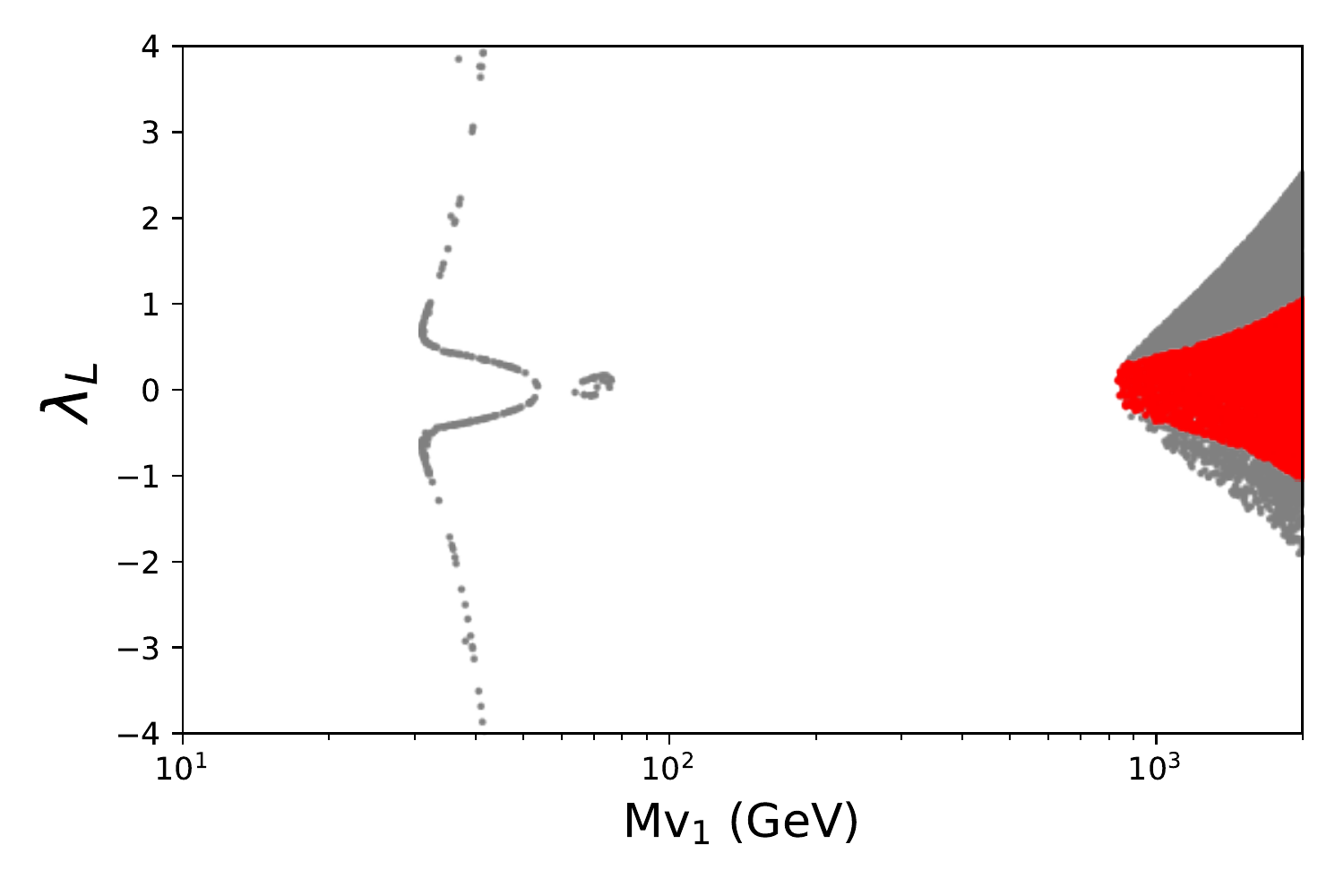}}%
{\includegraphics[width=0.50\textwidth]{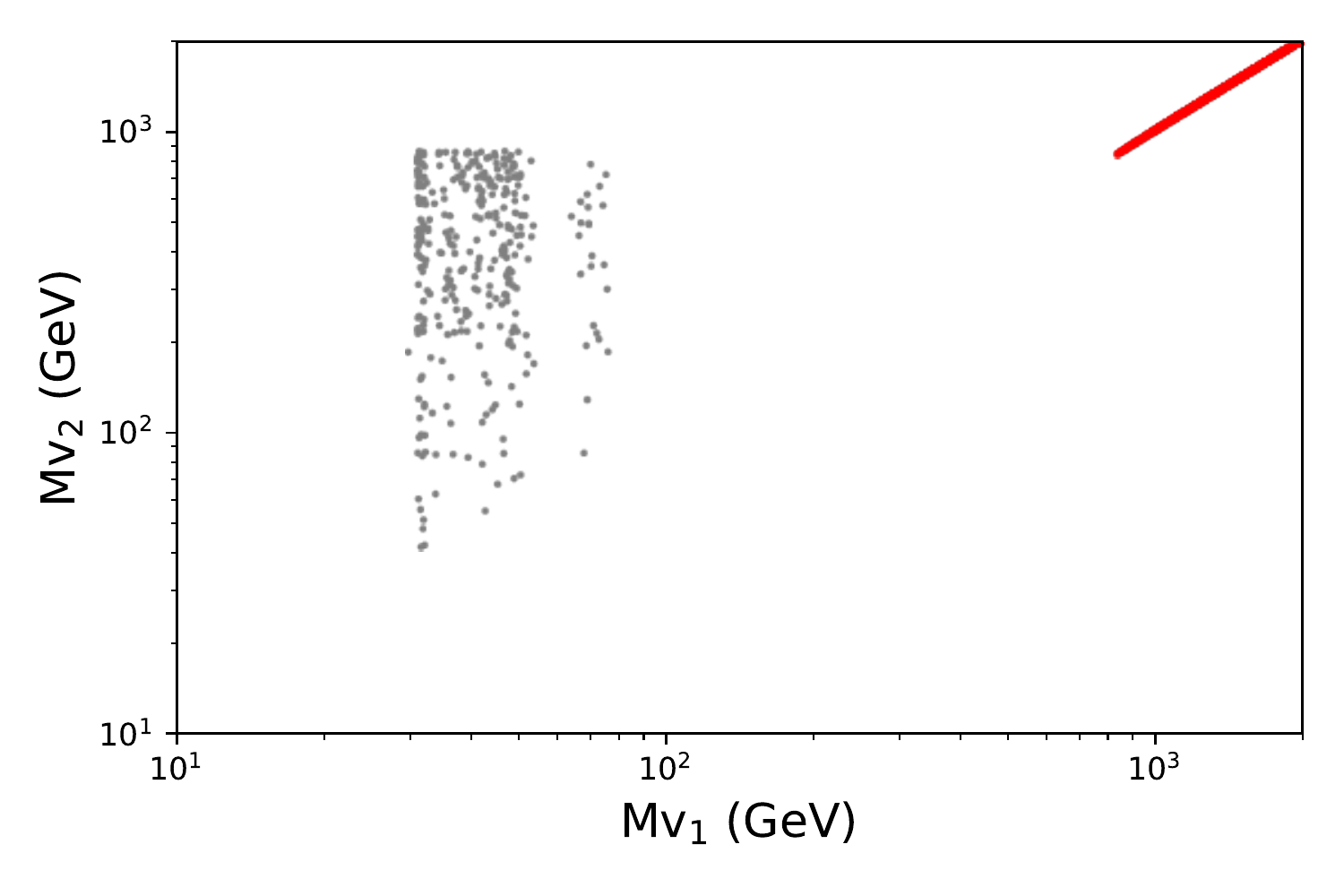}}%
\vskip -0.5cm\hspace*{-3cm}(a)\hspace*{0.48\textwidth}(b)
\caption{\footnotesize 2-D projections of the 4-dimensional parameter space in two different planes: ($\Mv[1],\lamL$) plane for Fig(a) and ($\Mv[1],\Mv[2]$) plane for Fig(b). We show all the points where the model fulfill the PLANCK limit, but the gray points are constrained by experiments and the red ones survive all the restrictions.} \label{fig:full-scan-sat}
\end{figure}

In the previous paragraphs, we considered experimental and theoretical constraints in our parameter space but we maintained the assumption that our DM candidates contributes partially to the DM budget, therefore we relaxed the lower limit of the measurements made by the PLANCK satellite. Here, we show how the model can completely explain the abundance of DM for some special region of the parameter space taking into account both upper and lower PLANCK limits at $1\sigma$ (\ref{PLANCK}). For that reason, in Figure.(\ref{fig:full-scan-sat}) we present a 2D projection of the 4-dimensional parameter space for the planes ($\Mv[1],\lamL$) and ($\Mv[1],\Mv[2]$), where we show all the points which can saturate the PLANCK limit but only the red points survived all the restrictions mentioned above.
\newpage

As we discussed earlier, there are two regions where the Vector DM reach the experimental limit. The first one happen in the low DM mass region between $35<\Mv[1]<80$ GeV. However this zone is complete exclude by the experimental constrains. The region of interest which survive after all the restrictions is located the high DM mass zone where $\Mv[1]\gtrsim 840$ GeV as we can see from Fig.(\ref{fig:full-scan-sat}). This result contrasts with the one found in references \cite{Mizukoshi:2010ky,Dong:2017zxo} where the dark vector can only explain partially the DM relic abundance. One of the most important features of this regions is the high level of degeneracy between the vector masses showed in the plane ($\Mv[1],\Mv[2]$) of Fig.(\ref{fig:full-scan-sat})(b) where the mass splitting do not exceeds $\Delta M<20$ GeV. Other works  where vector dark matter are presented can only explain partially the experimental 

\begin{figure}[htb]
\centering
{\includegraphics[width=0.51\textwidth]{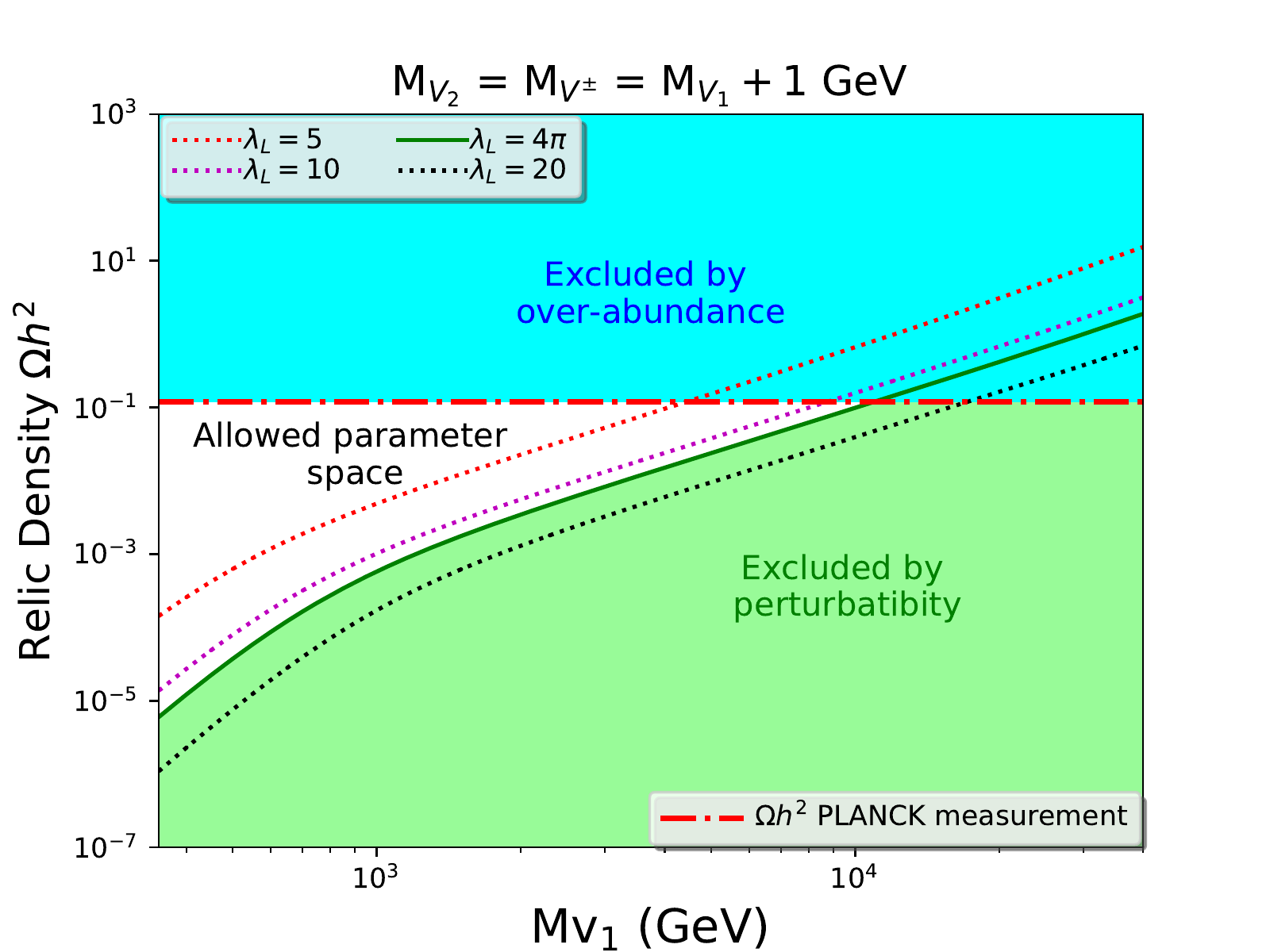}}%
{\includegraphics[width=0.51\textwidth]{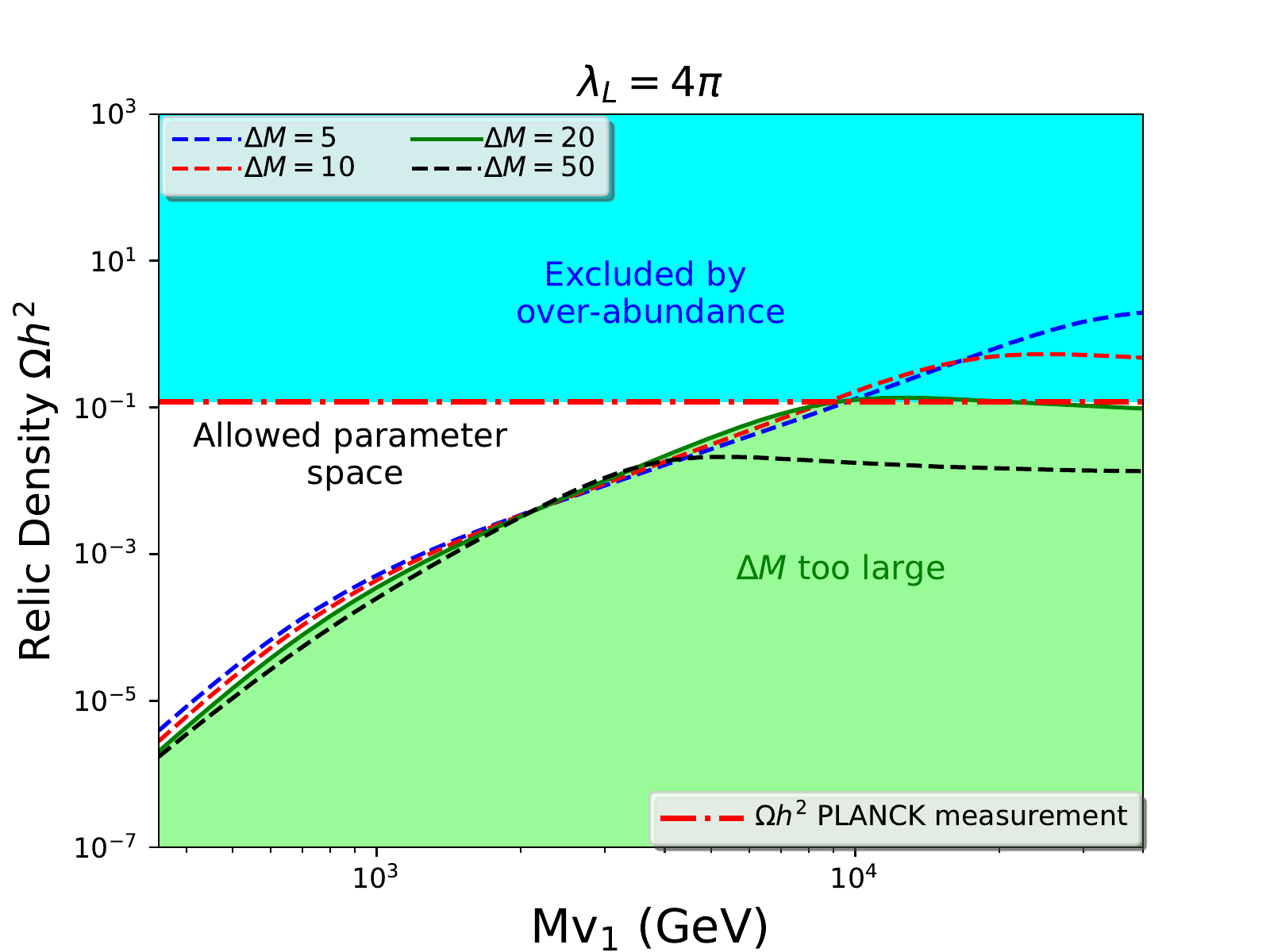}}%
\vskip -0.5cm\hspace*{-3cm}(a)\hspace*{0.48\textwidth}(b)
\caption{\footnotesize Closing of the parameter space at high values of $\Mv[1]$ and $\lamL$ in a quasi-degenerate scenario.} \label{fig:close}
\end{figure}

Despite the fact that direct detection experiment are less sensitivity in the zone of high DM mass, the XENON1T constraints are still able to exclude parameter space for $\lamL>0.3$ in this zone. As the value of $\Mv[1]$ increases and DD loses sensibility, the allowed region becomes bigger and higher values for $\lamL$ are allowed. This effect is appreciated as gray region for $\Mv[1]>840$ GeV in Fig.\ref{fig:full-scan-sat}(a).

As we increase the value of $\Mv[1]$ in this scenario of high degeneracy, we can notice that $\lamL$ can take larger values. However, when $\Mv[1] \sim 10$ TeV we reach the maximum value for $\lamL$ allowed by the perturbability constraints (\ref{pert}). Now, with this value of $\lamL$, the difference of masses between DM and the other vectors can only reach up to 20 GeV, after that point the quartic couplings become too large making the effective DM annihilation cross section fall below the experimental value of PLANCK. This completely closes the parameter space of the model as we can see from Figure.(\ref{fig:close}).

%%%%%%%%%%%%%%%%%%%%%%%%%%%%%%%%%%%%%%%%%%%%%%
\subsection{Dark matter production at the LHC}
%%%%%%%%%%%%%%%%%%%%%%%%%%%%%%%%%%%%%%%%%%%%%%
The DM double production associated with either mono-jet $j$, mono-$Z$ or a mono-$H$ are signals expected to been seen at the LHC in the context of dark matter searches. Due to the similarities in the topology of these processes between our model and the well know inert-two-Higgs-doublet-model (i2HDM) \cite{Barbieri:2006dq}\cite{Belyaev:2016lok}, we compare the parton level distribution cross section and the missing transverse energy shape in mono-X ($j,Z,H$) processes\footnote{Detailed analysis of DM production at LHC considering theses processes in the i2HDM see \cite{Belyaev:2016lok}, and a more fine analysis for mono-jet signature at the LHC see \cite{Belyaev:2018ext}.}. The calculations were made with \texttt{CalcHEP} package, using \texttt{NNPDF23\_lo\_as\_0130\_qed (proton)} as a parton distribution functions, and a generic transverse momentum cut of $100$ GeV on each of the SM particles. 

\begin{figure}[htb]
\centering
{\includegraphics[width=0.5\textwidth]{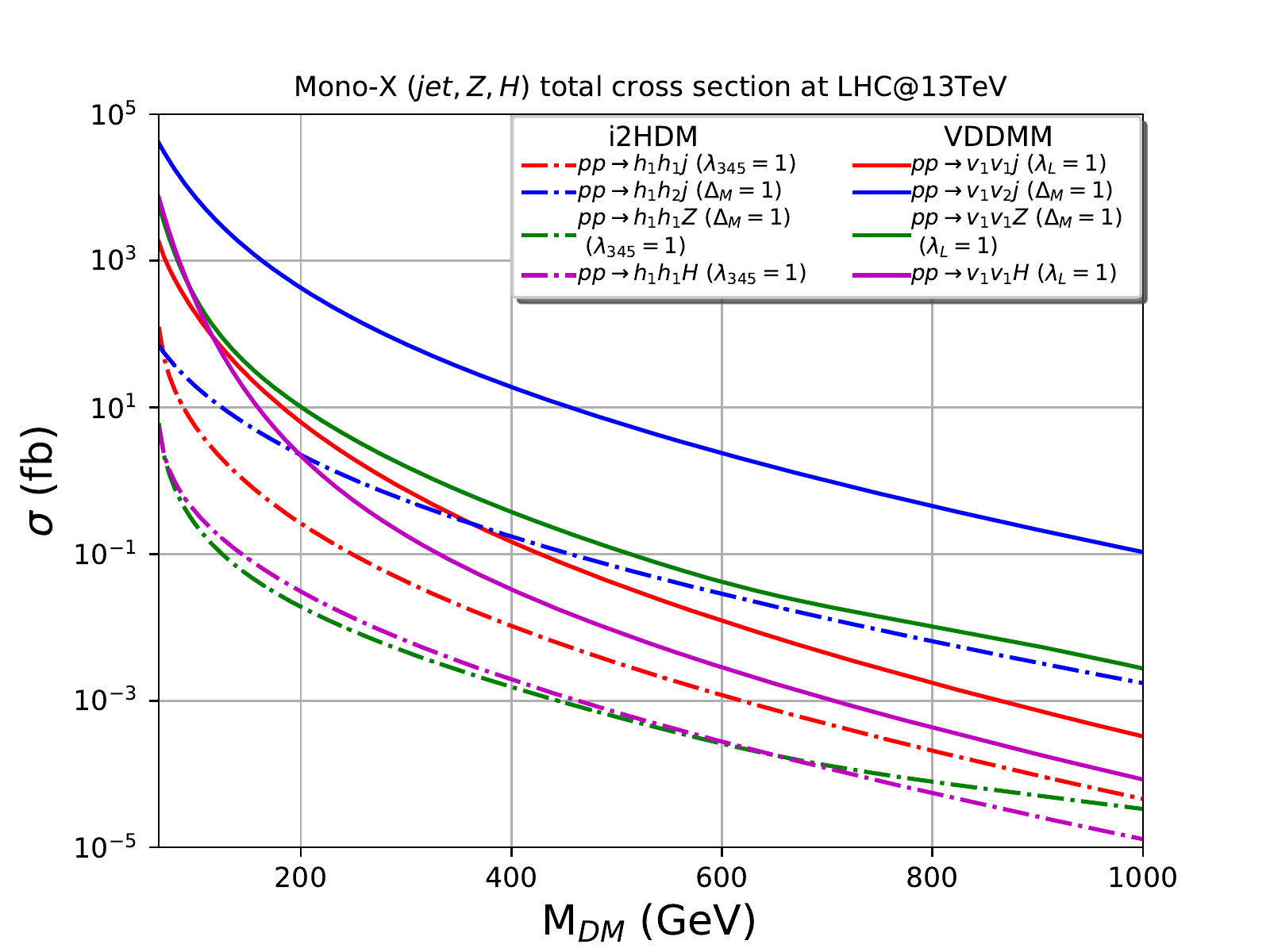}}%
\caption{\footnotesize Mono-X ($Z,j,H$) cross section as a function of the DM mass. The dashed lines correspond to the scalar case (i2HDM) and the continuous one to the vectorial one (DVDM).} \label{fig:cs}
\end{figure}

\begin{figure}[htb!]
\centering
\subfigure[]
{\includegraphics[width=0.5\textwidth]{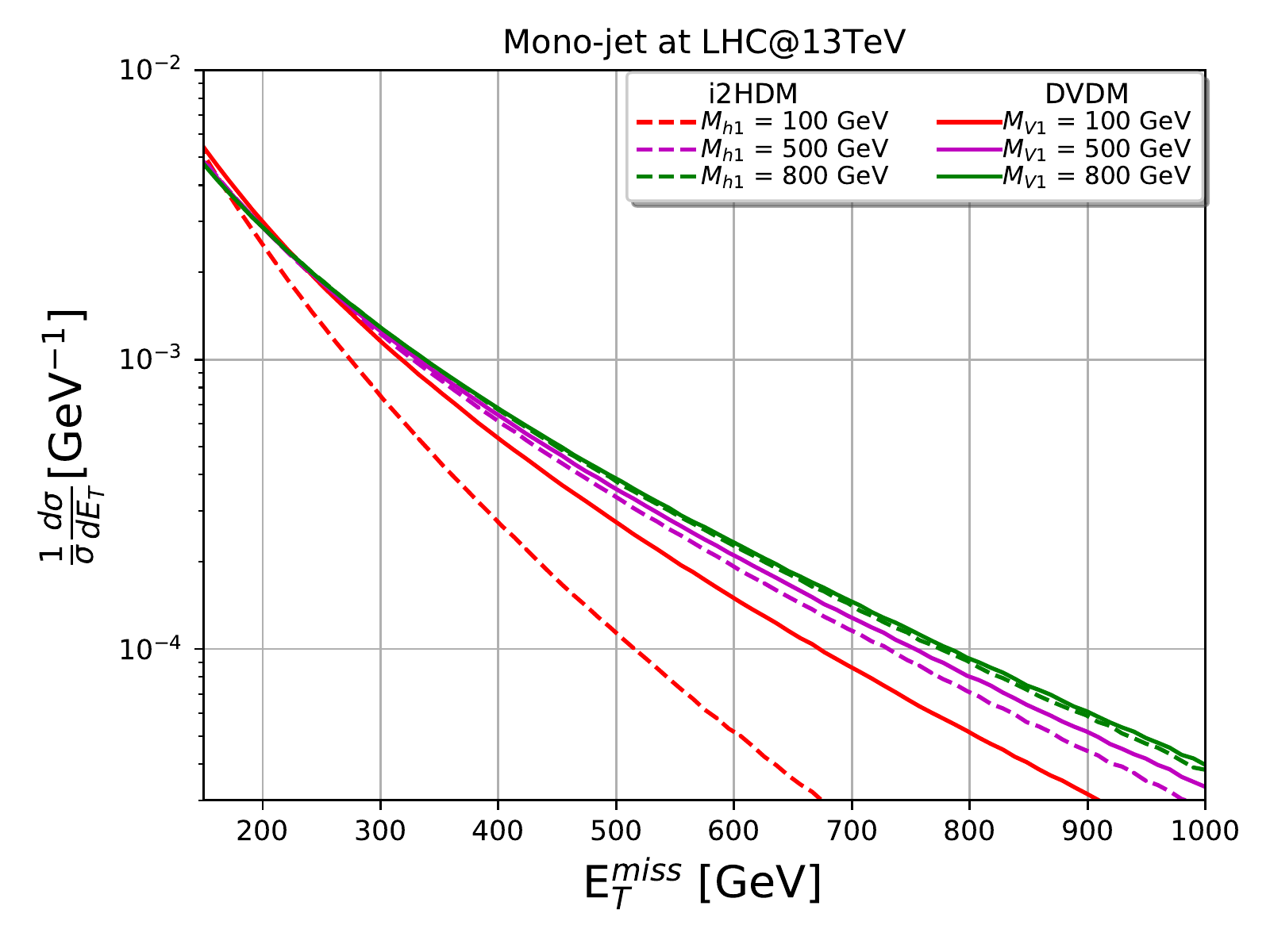}}%
\subfigure[]
{\includegraphics[width=0.5\textwidth]{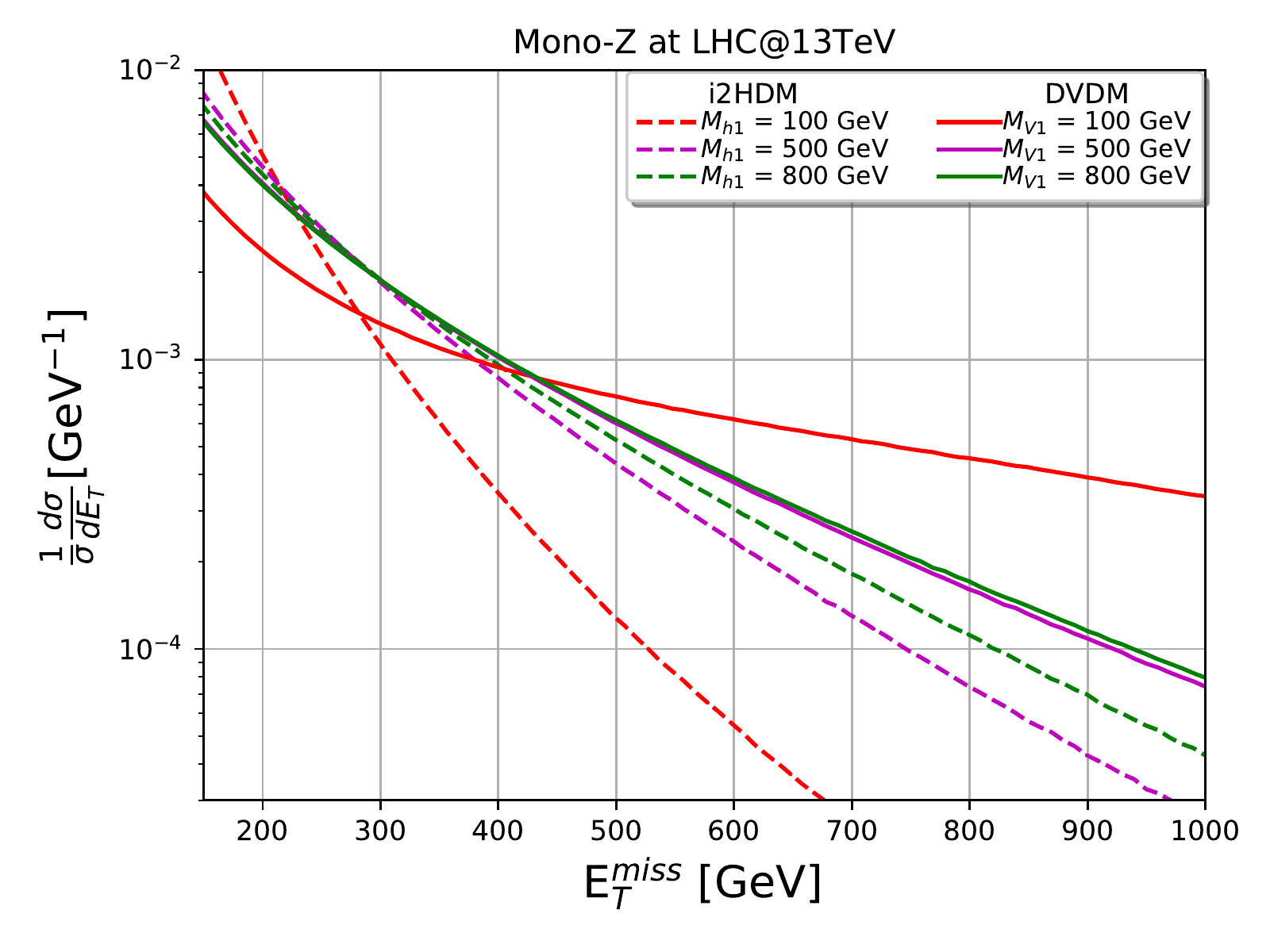}}%
\vspace{0.2cm}
%vskip -0.5cm\hspace*{-3cm}(a)\hspace*{0.48\textwidth}(b)
\subfigure[]
{\includegraphics[width=0.5\textwidth]{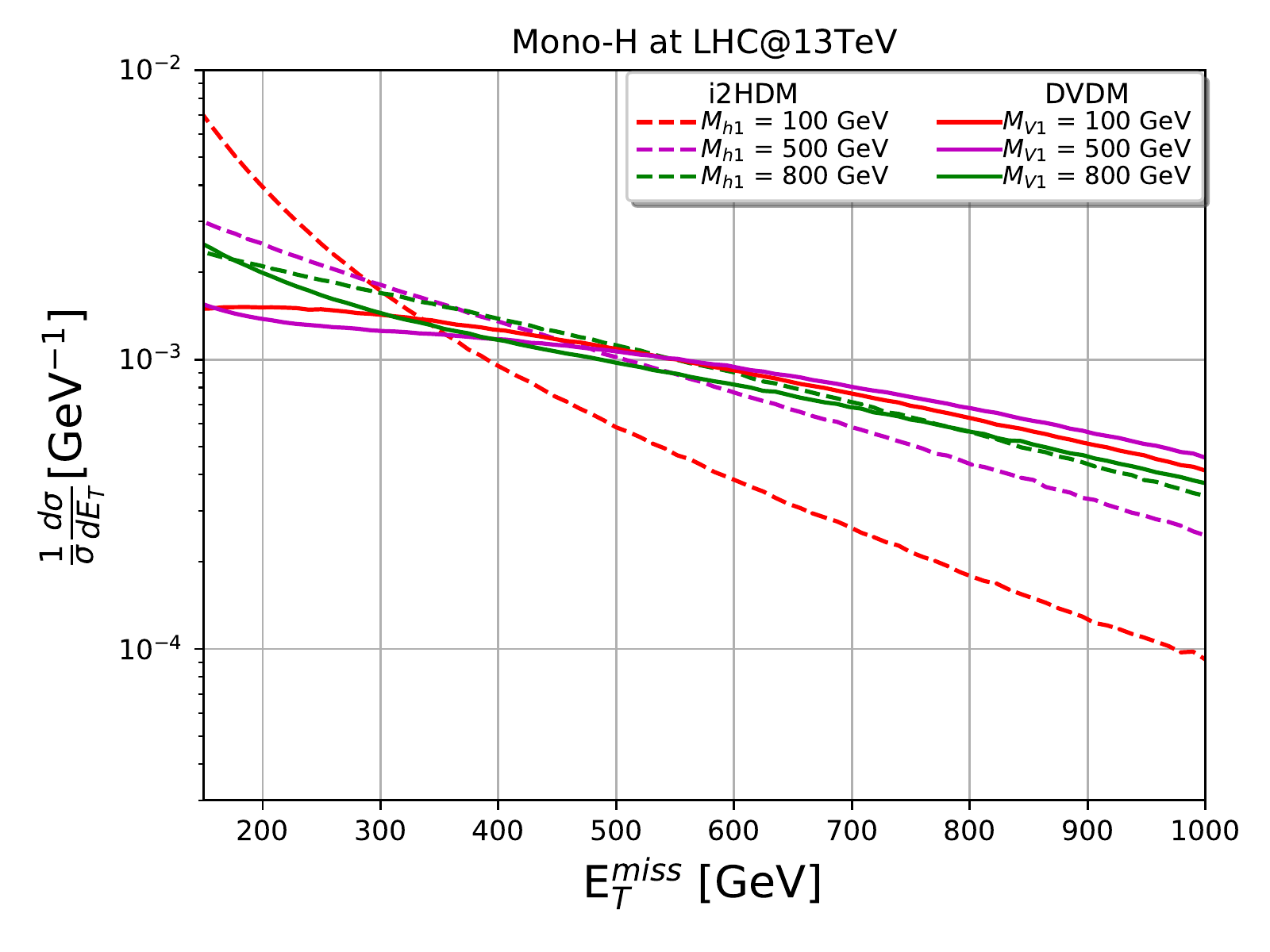}}%
\caption{\footnotesize Normalized differential missing transverse energy cross section for the processes $pp\rightarrow X+\slashed{E}$ ($X=j,Z,H$) in i2HDM and DVDM. The dashed lines correspond to the predictions of the i2HDM and the continuous one in the DVDM. All the plots contain different DM mass: 100, 500 and 800 GeV, in the quasi-degenerate case, i.e. $\Delta M=1$ GeV, $\lambda_{345}(\lambda_L)=0.1$, and at $\sqrt{s} = 13$ TeV LHC energies. A $p_{T,X} \geq 100$ GeV cut has been applied in all the plots.} \label{fig:monx}
\end{figure}

In Fig. \ref{fig:cs} we show the total cross section for the aforementioned processes as a function of the DM mass in both models for LHC@13TeV. The continuous lines correspond to the case of vector DM (DVDM), whereas the dashes lines to the scalar case (i2HDM). All the process consider the quasi degenerate mass scenario $\Delta M = 1$ and $\lambda_{345}(\lambda _L) = 1$. 

Because the topology of the Feynman diagrams in both models are exactly the same in all the processes studied here\footnote{The additional scalar states in the i2HDM are equivalent to the DVDM model, just do the replacement $V^{1,2}\rightarrow h_{1,2}$ and $V^\pm\rightarrow h^\pm$.}, the differences lies mainly in the spin of the final states. The  dependence of the cross section on the DM mass is similar in both cases. However the vector case is scaled up over the scalar one by, roughly, two orders of magnitude.
%The distribution cross section shapes in both models are very similar. However, as it is clear in Fig. \ref{fig:cs}, the vectorial cross section can reach up to two orders of magnitude bigger than the scalar one. 
This vector cross section enhancement is due to the fact that the longitudinal polarization of vectors scale as $\sim E/M_V$, implying that the production matrix element receive a significant enhancement in the region of phase space where the DM state is relativistic and either one or both particles are longitudinally polarized. 

\begin{figure}[htb!]
\centering
{\includegraphics[width=0.5\textwidth]{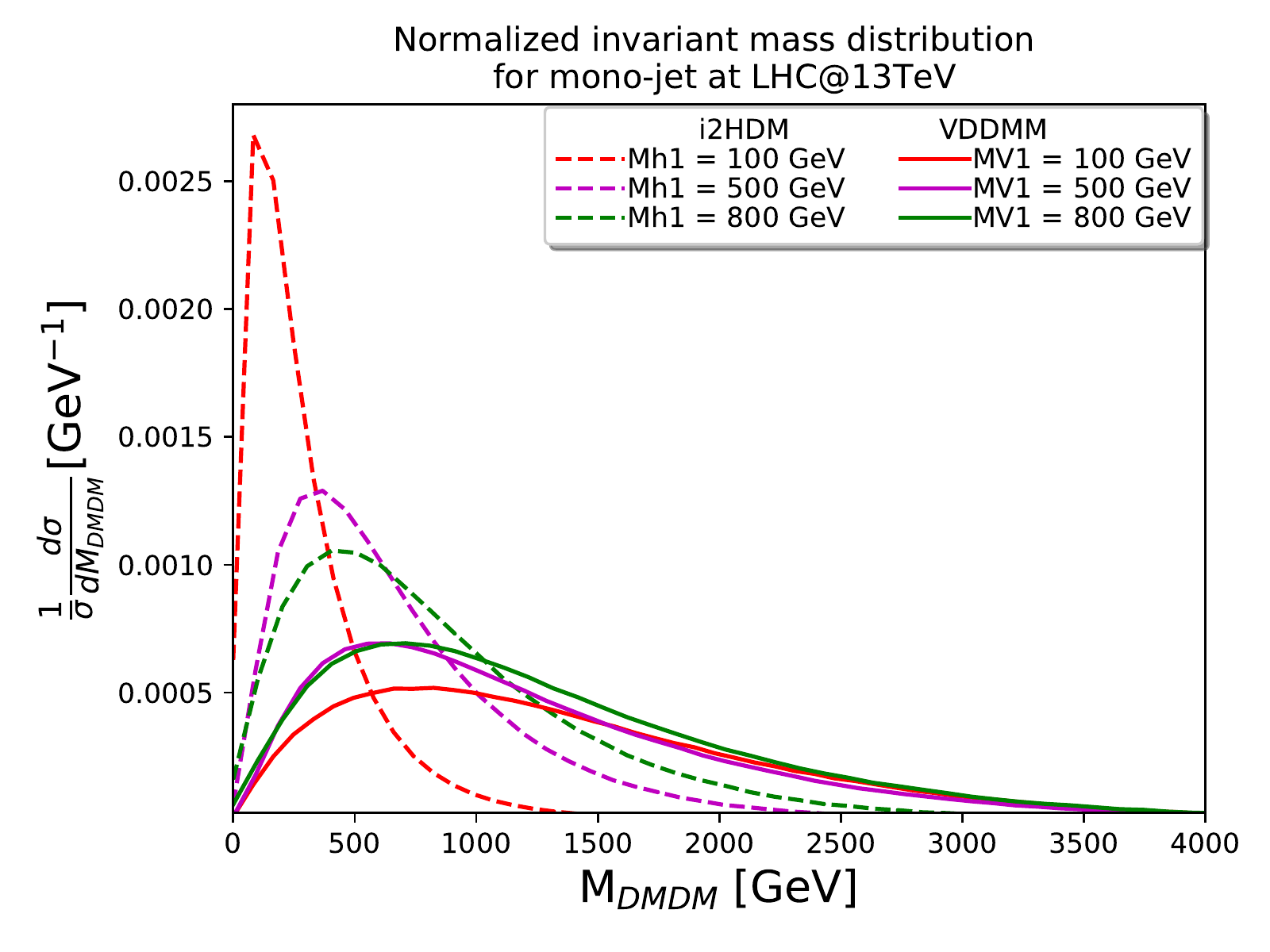}}%
\caption{\footnotesize Invariant mass of DM pair distributions normalized to unity for mono-jet in both i2HDM (dashed) and DVDM (continuous) at 13 TeV LHC energy. All the results are considering $\lambda_{345}(\lambda_L)=0.1$, in the quasi-degenerate case, i.e. $\Delta M=1$ GeV.} \label{fig:inv}
\end{figure}

On the other hand, in Fig. \ref{fig:monx} are shown the normalized missing transverse energy distribution cross section  of each one of the processes at parton level, considering the same mass splitting $\Delta M=1$ GeV and $\lambda_L = 0.1$.
%Each mono-X channel considered the sum of the leading diagrams, in which the missing transverse energy is given by $X_1X_1$+$X_1X_2$+$X_2X_2$, where $X_1 = h_1(V^1)$ and $X_2 = h_2(V^2)$.
 In each channel, the distributions for the vector case are always flattened respect to the scalar ones. This behaviour is in agreement with the results presented in \cite{Belyaev:2016pxe}. Furthermore, the differences in the shapes are more notorious in the cases in which the new state masses are lower. Considering that mono-jet signals have the higher cross sections, we complement the analysis with the invariant mass distribution of the DM pairs. In Fig. \ref{fig:inv} we present $M_{\text{inv}}(DM,DM)$ distributions for the scalar and vector cases in the mono-jet case, again normalized to unity for $\sqrt{s}$ = 13 TeV LHC energies. From Fig. \ref{fig:inv}, one can see that the $M_{\text{inv}}(DM,DM)$ distributions are better separated for higer masses of scalars and vectors. The scalar distributions are closer to the point $M_{\text{inv}}(DM,DM) = 2M_{\text{DM}}$, whereas the vectorial ones distributions are broader.
 % However, unlike \cite{Belyaev:2018ext}, for the regime $M_{h_1} > M_H/2$ we get that $h_1h_1j$ is slightly less steeply falling than $h_1h_2j$, and the same behaviour for the vectors ($\textbf{CORROBORAR ESTO}$). 

\begin{table}[ht]
	\caption{\footnotesize Total cross section (fb) for $pp\rightarrow X + \slashed{E},$ ($X=j,Z,H$) with $\lambda_{345}(\lambda_L) = 0.1$, \texttt{NNPDF23\_lo\_as\_0130\_qed (proton)} as a PDF, and $p_T^X > 100$ GeV. Here, the missing energy is due to the production of $V_1V_1$, $V_1V_2$ and $V_2V_2$. The same for the scalar case.}
	\vspace{-0.5cm}
	\begin{center}
		\begin{tabular}{|l|c|c|c|c|c|c|}
		\hline
		Model & \multicolumn{3}{|c|}{\textbf{i2HDM}} & \multicolumn{3}{|c|}{\textbf{DVDM}}\\
        \hline\hline\small
			Mass (GeV) & 100 & 500 & 800 & 100 & 500 & 800 \\ \hline \hline
			Mono-$j$ & $1.9\times 10^1$ & $6.6\times 10^{-2}$ & $6.4\times 10^{-3}$ & $7.3 \times 10^{3}$ & $6.2$ & $4.5\times 10^{-1}$  \\ \hline
			Mono-$Z$ & $3.7\times 10^{-1}$ & $2.9\times 10^{-3}$ & $3.1\times 10^{-4}$  & $7.3\times 10^{2}$ & $4.1\times 10^{-1}$ & $2.8\times 10^{-2}$ \\ \hline
			Mono-$H$ & $1.0\times 10^{-2}$ & $2.0\times 10^{-5}$ & $2.2\times 10^{-6}$ & $3.0\times 10^{2}$ & $3.1\times 10^{-3}$ &  $9.7\times 10^{-6}$  \\ \hline	
		\end{tabular} 
	\end{center}
	\label{csinv}
\end{table}

%%%%%%%%%%%%%%%%%%%%%%%%%%%%%%%%
\section{Perturbative unitarity}\label{pu}
%%%%%%%%%%%%%%%%%%%%%%%%%%%%%%%%

Having shown that our model can provide a viable Dark Matter candidate, we want to discuss the validity range of our effective approach. The main theoretical challenge faced by our construction is the eventual violation of perturbative  unitarity introduced by the new massive vector states. To this aim, we study the amplitudes, in the high energy regime, of representative and potentially problematic processes like $hV^1\rightarrow hV^1$ and $ZV^\pm\rightarrow ZV^\pm$ .

\begin{figure}[htb]
\centering
{\includegraphics[width=0.51\textwidth]{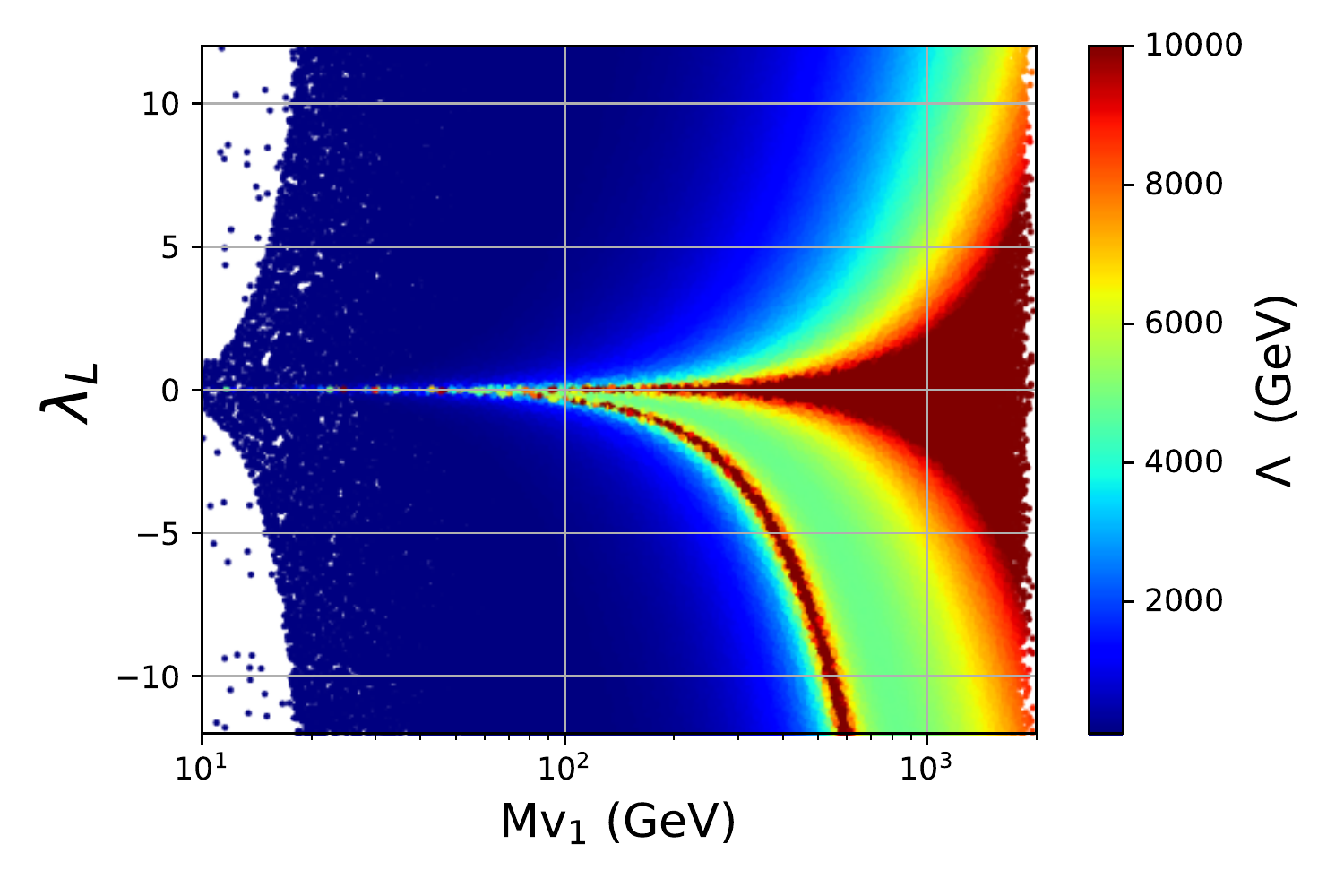}}%
{\includegraphics[width=0.51\textwidth]{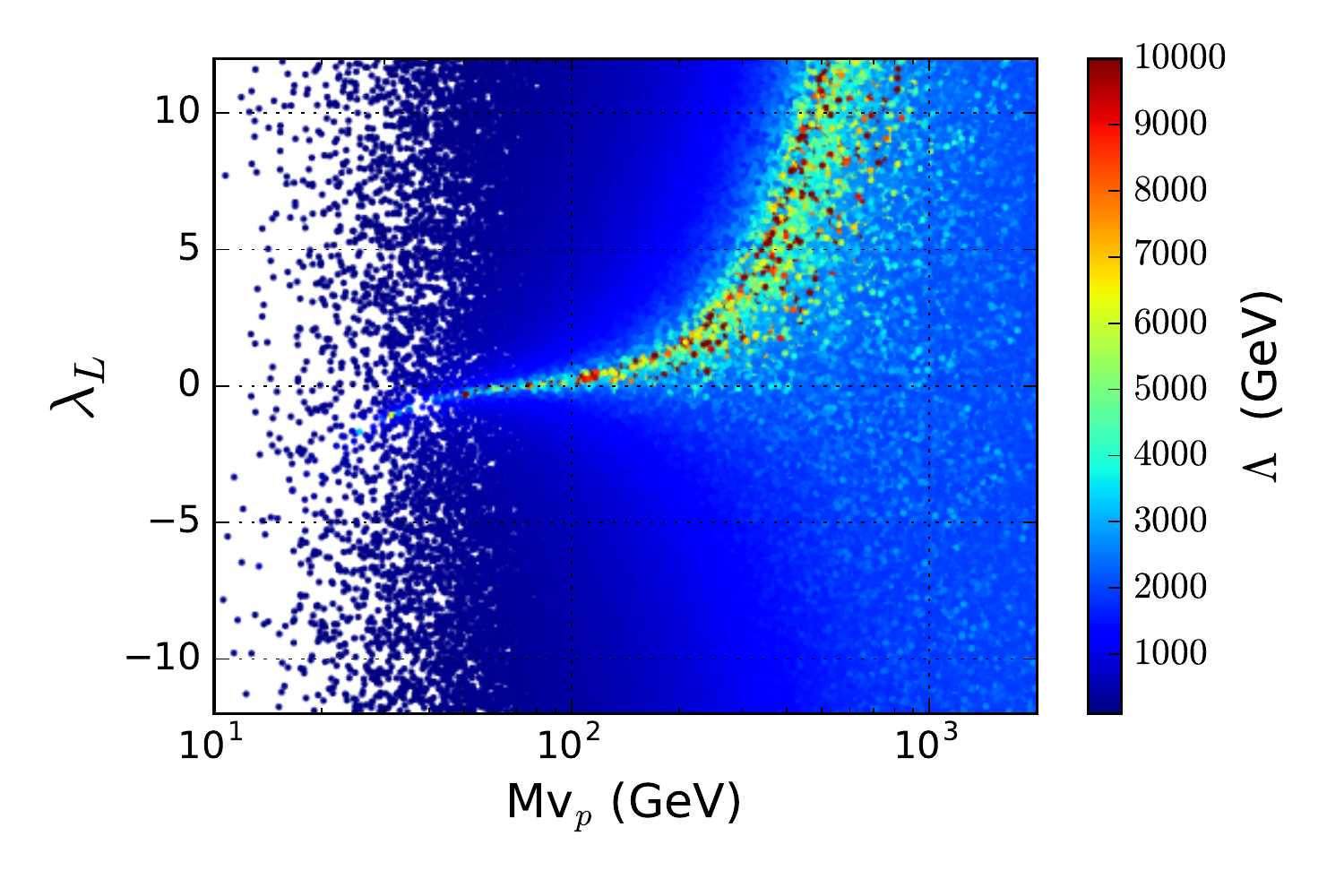}}%
\vskip -0.5cm\hspace*{-3cm}(a)\hspace*{0.48\textwidth}(b)
\caption {\footnotesize a) Maximum energy-scale $\Lambda$ until the process $hV^1 \rightarrow hV^1$ starts to violate perturbative unitarity. b) Maximum energy-scale $\Lambda$ until the process $ZV^\pm\rightarrow ZV^\pm$ starts to violate perturbative unitarity.} \label{pup}
\end{figure}

In Fig.\ref{pup}(a) is shown the maximum energy scale at which the process $hV^1 \rightarrow hV^1$ is valid until pertrubative unitarity start to be violated\footnote{The explicit expressions of the partial waves are in the Appendix.}. As $\lambda_L$ gets smaller the bigger is the scale energy before the breaking of pertrubative unitarity. Additionally, the bound on the energy gets relax as $M_{V^1}$ raises too. For values of $M_{V^1}$ below 100 GeV the scale of unitarity violation is mostly constant and of the order of a few TeV s, whereas for higher masses the dependence on $\lambda_L$ start to grow, making our model consistent at scales as high as 10 TeV for small values of $\lambda_L$. Therefore, from the point of view of unitarity, our construction is perfectly safe for masses of the DM candidate above 200 GeV specially when $\lambda_L$ to is small. We want to remark that phenomenologically interesting region of the space parameter, where our DM candidate saturate the relic density, belongs to the unitarity safe zone. 

 In Fig.\ref{pup}(b) is shown the maximum energy scale in the plane $(\lambda_L,M_{V^\pm})$ at which the process $Z_LV^\pm_L\rightarrow Z_LV^\pm_L$ is valid until perturbative unitarity is violated. At masses near 100 GeV and $\lambda_L$ close to zero, the maximum energy values allowed by pertrubative unitarity rises easily above $5$ TeV. on the other hand, for values of $M_{V^\pm}$ near 1 TeV, the scale of unitarity violation is of the order of $3$ TeV.

 These results are consistent with unitarity analysis of some vector dark matter models \cite{Lebedev:2011iq}\cite{Belyaev:2018xpf} and suggest that our effective model must meet an ultraviolet completion at a scale between $3$ and $10$  TeV. For instance, one of the simplest ways to restore unitarity is to embed our model into a larger gauge symmetry spontaneously broken by a new scalar sector \cite{Kahlhoefer:2015bea}. In this sense, our model can be considered as a simplified model \cite{Abdallah:2015ter}, retaining just the lightest states predicted in this scenario, and pushing the required new states at scales above the vectorial ones.

\section{Conclusions} \label{Conclusions}
Unlike most of extensions to the Standard Model which consider new massive vector fields as singlets or triplets under $SU(2)_L$ gauge group, in this work we have explored a different possibility. The new vector degrees of freedom enter into the SM in the fundamental representation of $SU(2)_L$, with hypercharge $Y=1/2$. Unlike vector triplet case, our model accept a potential composed of many terms coupling the new vector to the Higgs doublet with independent coupling constants. This feature makes the model more similar to the i2HDM than to models with vector triplets. Additionnaly, due to the quantum numbers assigned to the new vector, it is impossible to couple it to standard fermions through renormalizable operator. 
% in this representation the Higgs sector may couple to the new degrees of freedom with different coupling, and naturally forbids a renormalizable coupling among the SM fermions and the new states.
 %This kind of vectors has been motivated in different context, such as Higgs-Gauge unification, Composite-Higgs, and extra dimensions. 

The model acquires a $Z_2$ symmetry in the limit in which the only non-standard dimension three operator is eliminated. This choice is natural in the sense of t'Hooft and allows the neutral vector $V^1$ to be a good dark matter candidate.

%Although the model introduces six additional parameters: three physical masses of the vector particles ($M_{V^1},M_{V^2},M_{V^\pm}$), one coupling between DM and SM-Higgs ($\lambda_L$), and two quartic couplings of self-interaction among the new vector fields, ($\alpha_2$, $\alpha_3$). In our study we reduced the parameter space to four of them considering that the two last have a poor sensitivity in the study of DM phenomenology, and we imposed pertrubative condition on each of the coupling constants. We implemented the model in \texttt{CalcHEP} and \texttt{micrOMEGAs} framework in order to do most of the computations.

We have performed a detailed analysis constraining the model through LEP and LHC data, DM relic density and direct DM detection. We found that the main experimental constrains are imposed by recent measurement of $H\rightarrow \gamma \gamma$ (mainly when the $V^1$ is light) and data on   direct search of DM obtained by XENON1T.
 
%LEP limits put constraints on vector physical masses, and also on certain combinations of them. Higgs diphoton decay can have sizable contributions at one-loop. ATLAS measurements put severe constraints on the model, excluding most of the low DM masses ($M_{V^1}\leq M_H/2$), because in this region the channel $H\rightarrow V^1V^1$ is opened and then the signal strength $\mu^{\gamma\gamma}$ gets suppressed. Furthermore, for $\Mvc \gtrsim 400$, $\lambda_2$ parameter gets highly constrained to be small, constraining directly the possibles values of both $\lambda_L$ and DM mass. For very high masses, above 1 TeV, diphoton constraints relax the possibles values of $\lambda_2$. Invisible Higgs decay put very high constraints on vector with masses $M_H/2$. 

After impossing all the experimental constraints, we found that for a range of masses between $840 \leq \Mv[1] \leq 10^4$ GeV in the highly degenerate case where $\Delta M < 20$ GeV the lightest neutral component of the doublet can reach the relic density measurements \ref{PLANCK}, surviving all the experimental constraints. This contrast with other electroweak vector multiplets models, where the saturation value for DM is above the TeV scale (see e.g. \cite{Belyaev:2018xpf, Maru:2018ocf}), or other models where the dark vector component never reach the DM budget (see e.g. \cite{Mizukoshi:2010ky,Dong:2017zxo}). Furthermore, if we relax the lower PLANCK limit \ref{PLANCK} allowing additional sources of dark matter, there is an important sector of the parameter space for $\Mv[1]\gtrsim 100$ GeV that it is still not possible to rule out with the current experiments.

At this point, we want to dedicate some sentences to compare our construction to the recently proposed Minimal Vector Dark Mater model (MVDM) \cite{Belyaev:2018xpf}. In both models, the dark matter candidate is a component of a vector field transforming a non-trivial representation of 
$SU(2)_L$: the adjoint representation in the case of MVDM and the fundamental one in our case. The difference in representations makes an abysmal separation between the two models. The most evident one is related to the number of new vector states ($3$ for the MVDM and $4$ in our case). But more important is what happen with the potential in the Higgs--massive-vector sector. In the MVDM this sector is extremely simple, contributing with only one term to the Lagrangian and only one of the two free parameters of the model. In our case, the scalar-vector potential is richer with three  free parameters. This is, in part, the origin of the different ultra-violet behavior reflected in the scale of unitarity violation which is systematically larger in the MVDM. In fact, the structure of the potential of our model makes it more closely related to the i2HDM than to the MVDM making harder to differentiate our model from the former than from the latter.

%In contrast with other electroweak vector multiplets, it is possible to have vector DM which saturates below the TeV scale, passing most of the LHC and experimental constraints (en MVDM encuentran vectores de masa altísima, en 1710.06951 arugumentan que sus vectores nunca sautran, o en 1803.01274 ellos obtienen masas de 1.5 TeV).

In view of the similarities between i2HDM with our model, we compared the parton level cross section and the normalized missing energy differential cross section for mono-$X$(jet,$Z$,$H$). Mono-X cross section get enhanced in the vectorial case due to their growing energy behaviour of their final state longitudinal polarization. The shapes of the distribution of missing energy results to be  flatter in the vectorial cases. This feature may help to distinguish between our model and the i2HDM. 

Finally, as a complement to this work, we have shown some results of perturbative unitarity bounds on some scattering amplitudes involving the new states. Our analysis suggest that our effetive approach needs an ultraviolet completion at a scale of the order of $3$ to $10$ TeV.

%This analysis give us an insight on the validity of perturbation theory in the model at arbitrarily high energies. We have shown that elastic process contrained in a different regime of masses the allowed energies before break pertrubative unitarity. The growing energy behavior can be relaxing making the UV-completion of the model introducing necessarily new scalar degrees of freedom that cancel the dangerous amplitudes, such as the role that the Higgs boson plays in the SM. The UV-completion is beyond the present work, but just a few comments can be done in that context.

\section*{Acknowledgements}	
We would like to thank Marcela Gonz\'alez for
contributions at early stages of this work, and also to Alexander Belyaev, Eduardo Pontón and Sebastián Norero for valuable discussions. BD was supported partially by Conicyt Becas Chile and DGIIP-UTFSM. AZ was supported in part by Conicyt (Chile) grants PIA/ACT-1406 and PIA/Basal FB0821, and by Fondecyt (Chile) grant 1160423. FR was supported by Conicyt Becas Chile Postdoctorado grant 74180065. AZ is very thankful to the developers of MAXIMA \cite{maxima} and the
package Dirac2 \cite{Dirac2} 
. These softwares were used in parts
of this work.
\appendix
\section{More on partial waves amplitudes} \label{app:A}
In this appendix we complement the results of the analysis presented in section \ref{pu} with more details. Additionally, we make some comments about perturbative unitarity on self-interaction amplitudes ($VV\rightarrow VV$, for $V=V^1,V^2$ or $V^\pm$).
\begin{figure}[htb!]
\centering
\subfigure[]
{\includegraphics[width=0.45\textwidth]{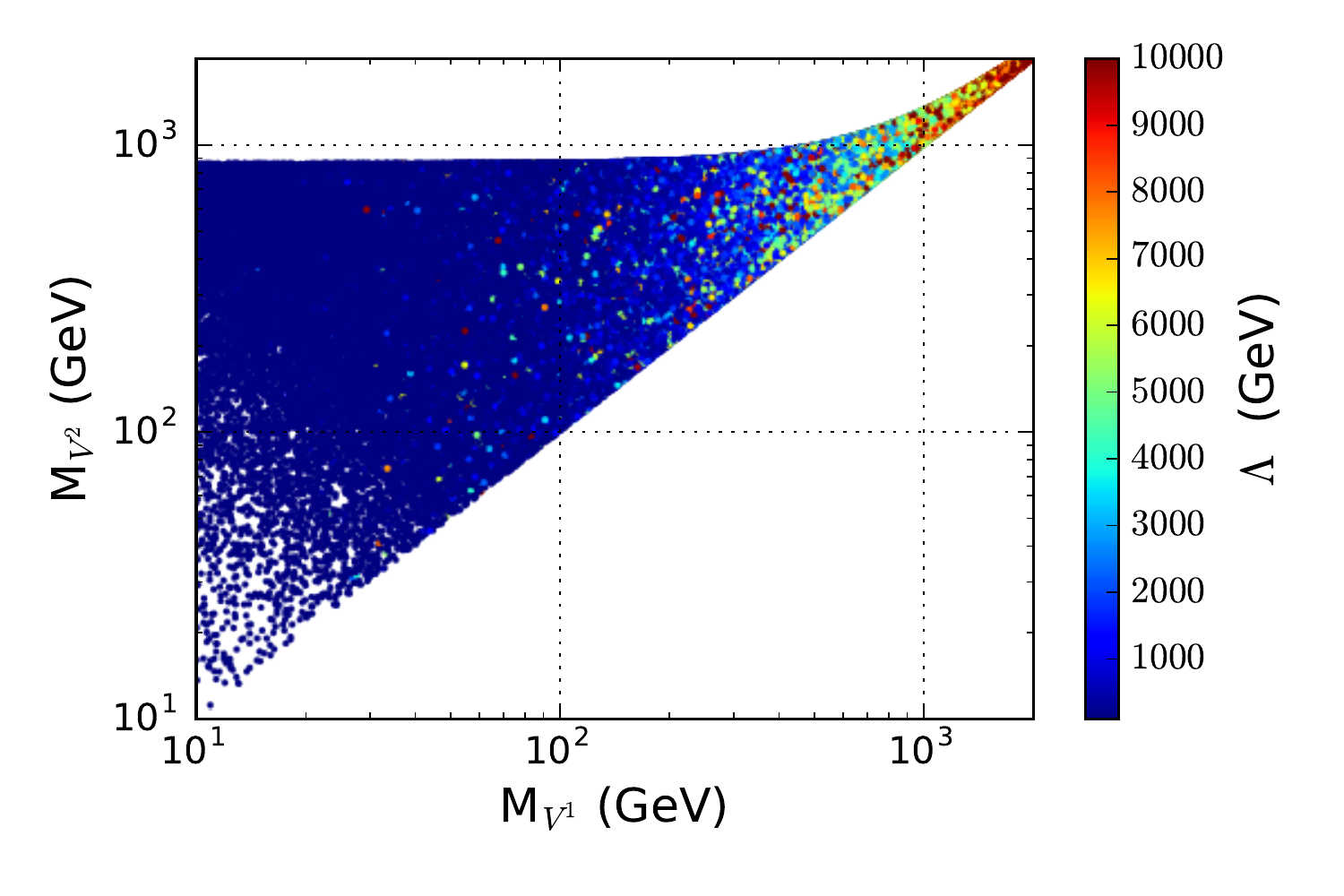}}
\subfigure[]
{\includegraphics[width=0.45\textwidth]{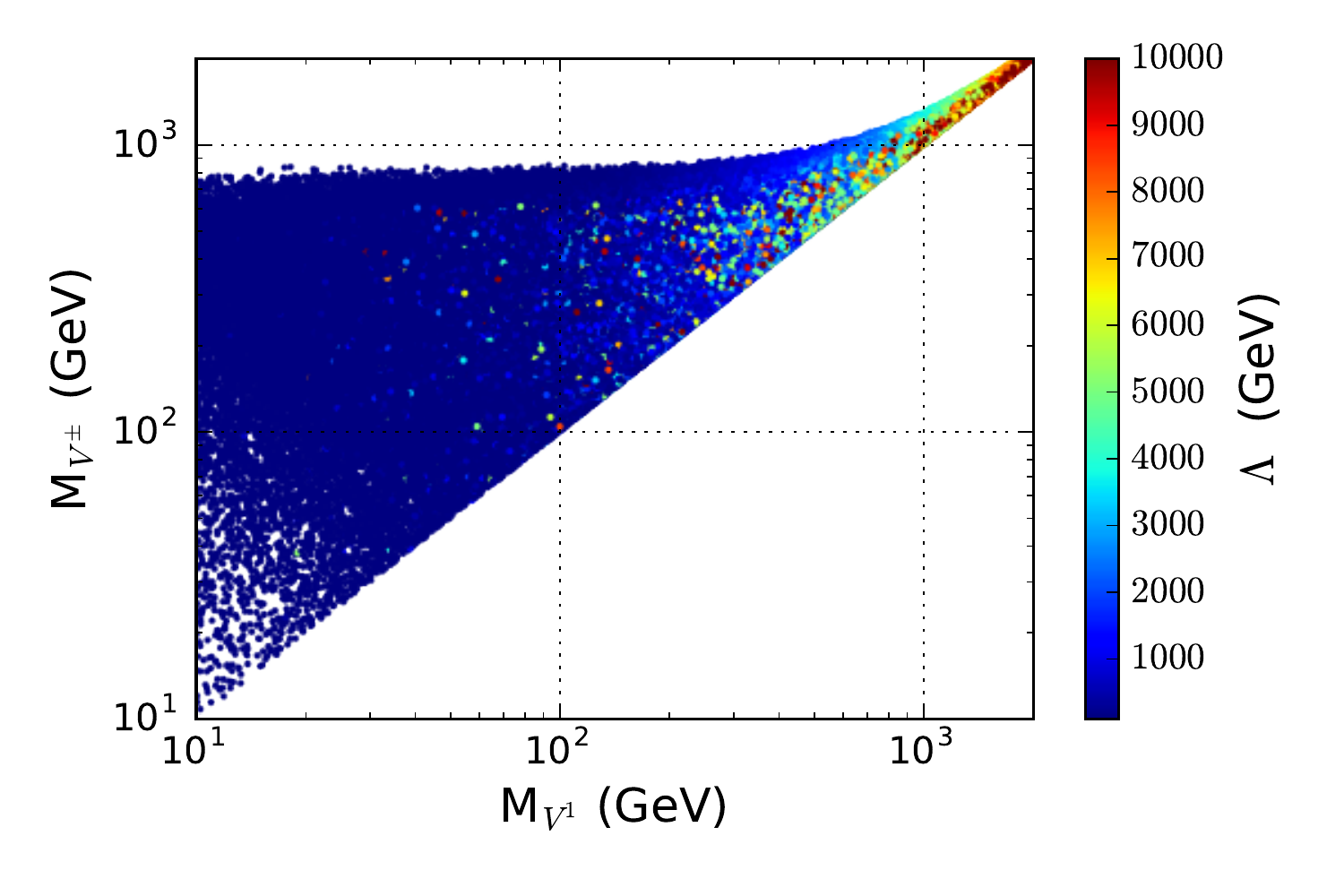}}
\vspace{0.2cm}
\subfigure[]
{\includegraphics[width=0.45\textwidth]{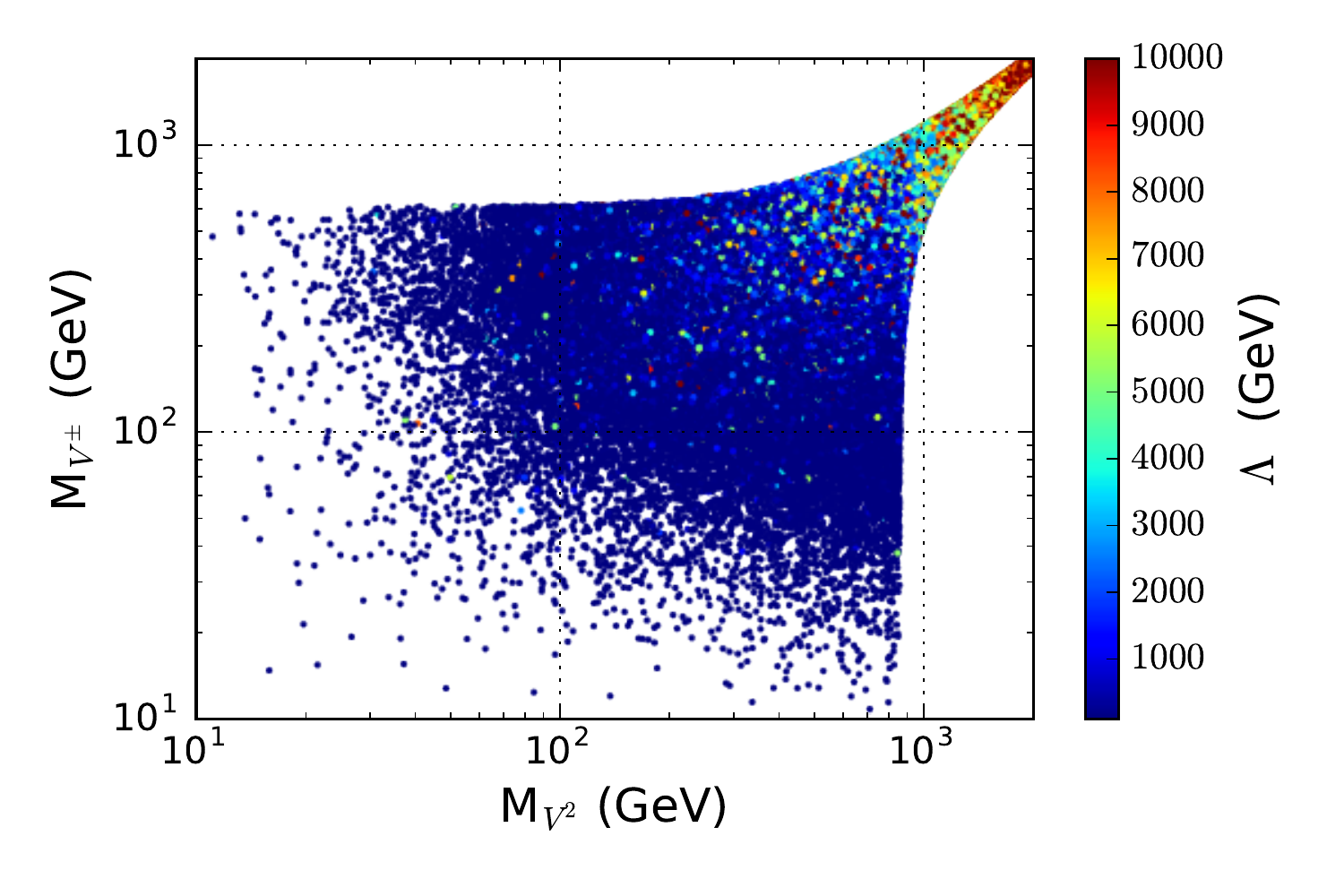}}
\caption{Maximum allowed energy $\Lambda$ by perturbative unitarity bounds on $hV^1 \rightarrow hV^1$ amplitude. The plots are projected in the planes ($M_{V^1},M_{V^2}$)(a), ($M_{V^1},M_{V^\pm}$)(b) and ($M_{V^2},M_{V^\pm}$)(c).} \label{fig:hVAP}
\end{figure}

In Fig. \ref{fig:hVAP} is shown the value of the scale of perturbative unitarity violation  ($\Lambda$) for the process $hV^1\rightarrow hV^1$ in the planes $(M_{V^1},M_{V^2})$, $(M_{V^1},M_{V^\pm})$ and $(M_{V^2},M_{V^\pm})$, respectively. In table(\ref{tab_partial}) we ressume the zero partial wave for the three possibles elastic scattering of this type. In concordance with the information given by Fig.\ref{pup}(a), for lower masses ($\lesssim 200$ GeV), the values of $\Lambda$ are located around the TeV energy scale for most of the  masses combinations allowed by experimental constrains. For higher masses, $\Lambda$ stars to grow for most of possible combination of masses, and there is a slightly raising in the energy as the degeneracy among the three states becomes similar.  

\begin{figure}[htb]
\centering
\subfigure[]
{\includegraphics[width=0.45\textwidth]{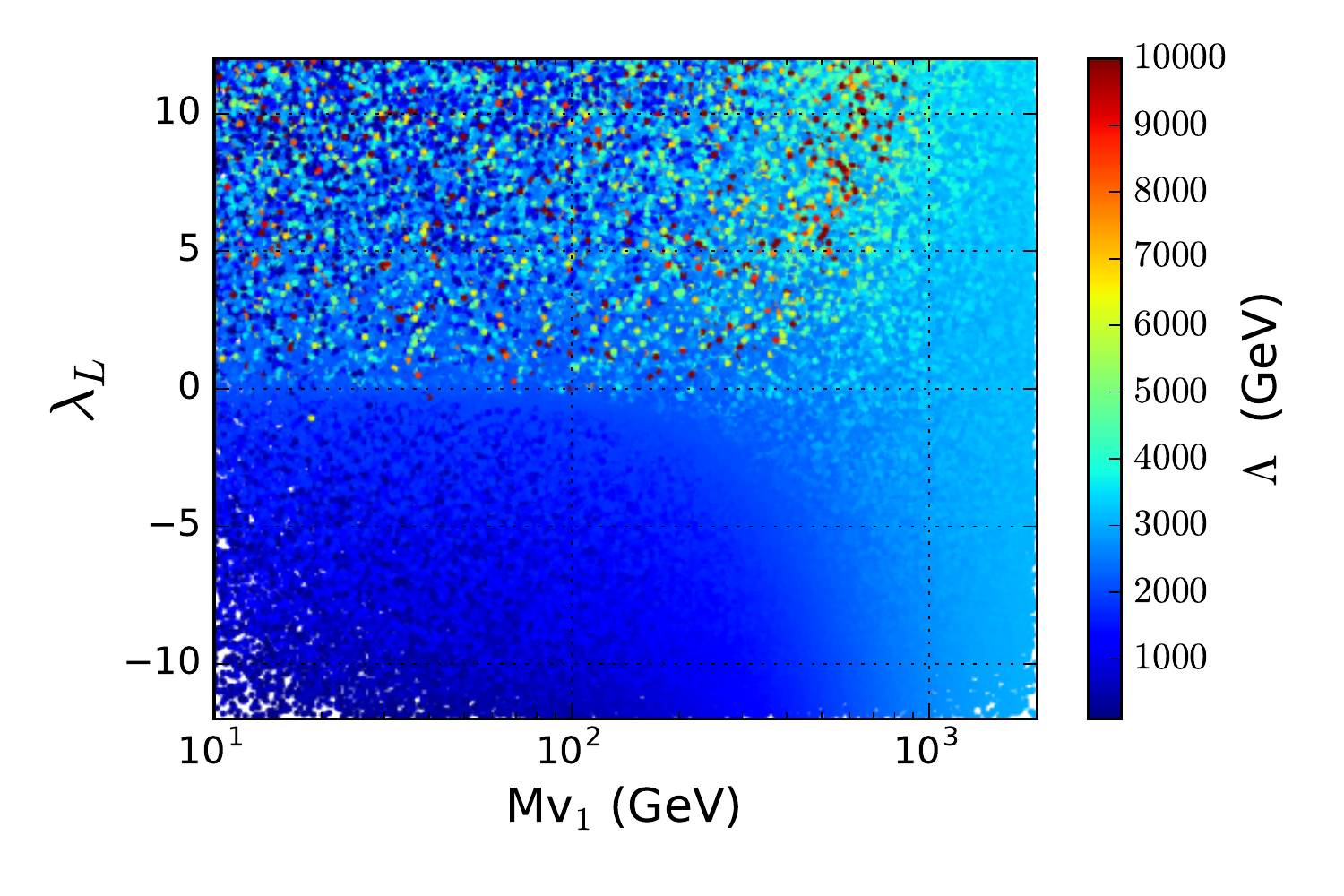}}
\subfigure[]
{\includegraphics[width=0.45\textwidth]{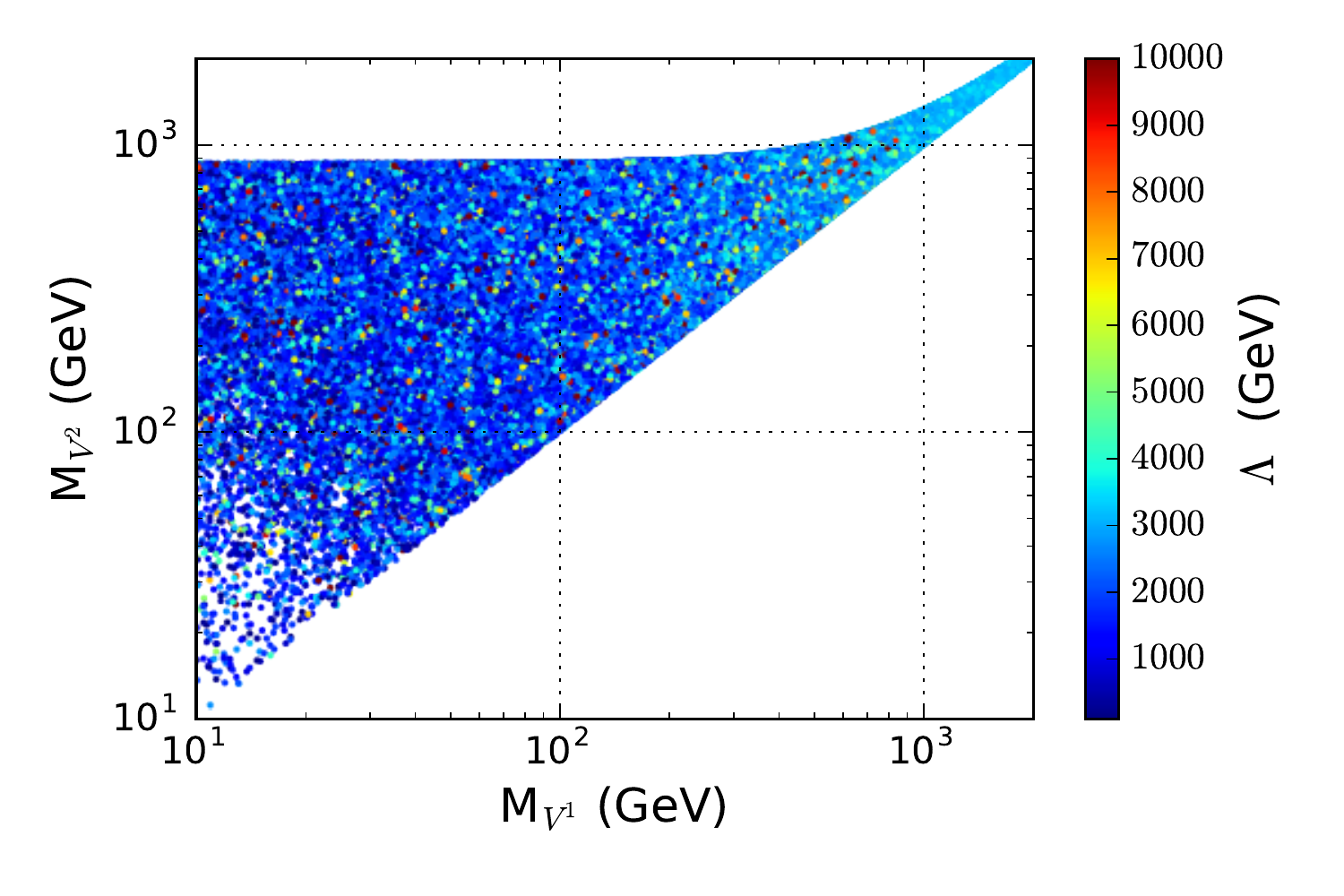}}
\vspace{0.2cm}
\subfigure[]
{\includegraphics[width=0.45\textwidth]{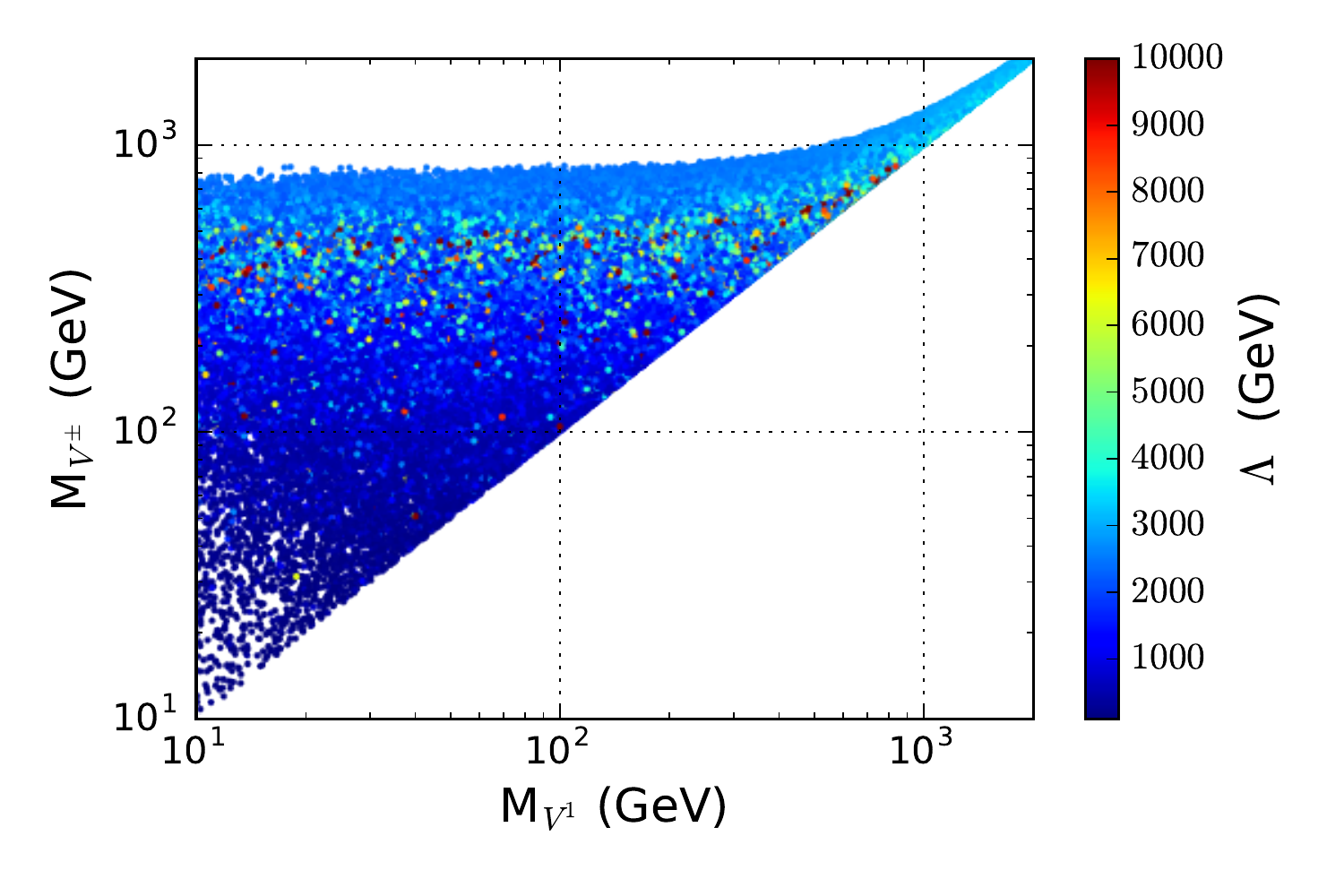}}
\subfigure[]
{\includegraphics[width=0.45\textwidth]{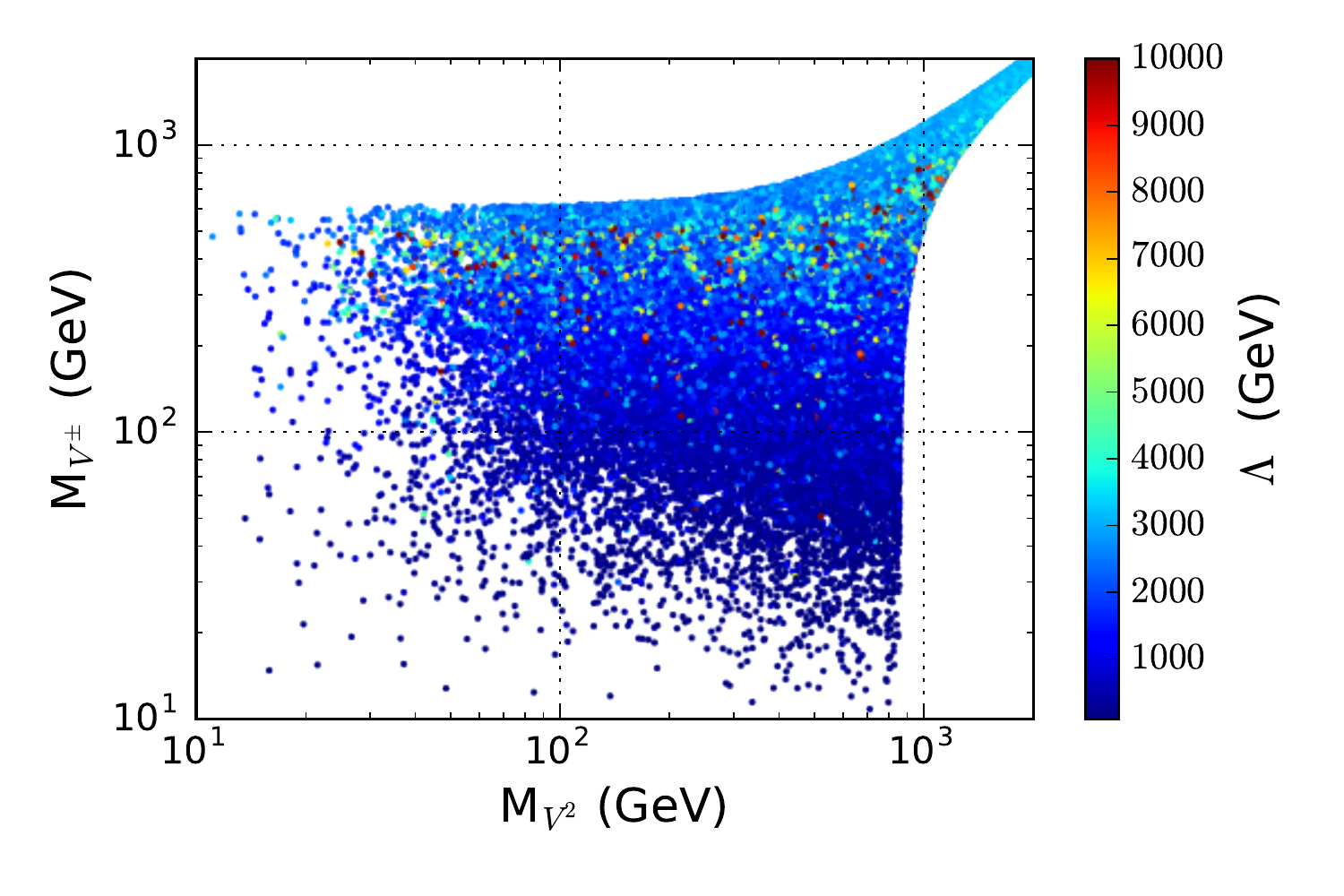}}
\caption{Maximum allowed energy $\Lambda$ from perturbative unitarity bounds on the processes $ZV^\pm \rightarrow ZV^\pm$ in the planes (a)($M_{V^1},M_{V^2}$), (b)($M_{V^1},M_{V^\pm}$) and (c)($M_{V^2},M_{V^\pm}$).} \label{fig:MVMV}
\end{figure}

On the other hand, in Fig. \ref{fig:MVMV} we present different plots showing the values of $\Lambda$  for the process $ZV^\pm\rightarrow ZV^\pm$. In this case, the degenarancy of the states do not show any raising in the maximum allowed energy value. According to what is shown in Fig.\ref{pup}(b), as the masses get near the TeV scale, $\Lambda$ gets a constant value near $4$ TeV, making this process more stringent for masses above $\sim 500$ GeV than the previous one with the Higgs involved. 

Finally, we make some comments about the $V_L+V_L\rightarrow V_L+V_L$ amplitudes, for $V=V^1,V^2$ and $V^\pm$. These processes may introduce strong constraints on the energy scale at which perturbative unitarity breaks down. For example, let us first consider the process $V^1_L+V^1_L\rightarrow V^1_L+V^1_L$ . Its zero partial wave is
\begin{eqnarray}
 a_0(s) = \frac{g^2(\alpha_2 + \alpha_3)\left(36M_{V^1}^4 - 24M_{V^1}^2s + 5s^2\right) + 18\lambda_L^2 M_H^2M_W^2}{96\pi g^2M_{V^1}^4},
\end{eqnarray}
where $\alpha_2$ and $\alpha_3$ are the self-couplings among the new states (see \ref{eq:fl}). The strong growing energy behaviour of the partial wave ($a_0 \sim E^4$) makes that perturbative unitarity breaks down at very low energies for typical masses of a few hundred GeV. For example, for $M_{V^1} = 100$ GeV, $\lambda_L=1$ and $\alpha_1=\alpha_2 = 1$, the breaking of perturbative unitarity is reached at energy scales less than 250 GeV. Interestingly, the growing energy behaviour dissapear when $\alpha_1 = -\alpha_2$. However, under this last condition, the amplitude $V^+_L+V^-_L\rightarrow V^+_L+V^-_L$ still grows with the energy as $E^4$:
\begin{eqnarray}
 a_0(s) &=& \frac{\frac{g^2(9(1-2c_w^2)^2M_W^2s - 4c_w^2s^2)}{c_w^4M_{V^\pm}^2} + \frac{48M_W^2\lambda_2 s}{g^2M_{V^\pm}^2} - 32(\alpha_2+\alpha_3)\left(\frac{2s^2}{M_{V^\pm}^2 - 9} \right)}{1536\pi M_{V^\pm}^2},
\end{eqnarray}
where $\lambda_2$ is a function of $\lambda_L$, $M_{V^1}$ and $M_{V^\pm}$ (see eq. \ref{eq:lam2}), and the lost of perturbative unitarity starts to be around 3 TeV. Therefore, it seems impossible to get rid of the growing energy behaviour with an arbitrarily choose of the free parameters. As we have pointed out in section \ref{pu}, a possible solution to this problem is to establish the model from a gauge theory in order to generate a gauge cancellation among the $s$- and $t$- channels and the contact graph \cite{Lee:1977eg}.

\begin{table}
\caption{Partial waves for $h V \rightarrow h V$ elastic tree level scatterings processes. Each of the three processes contain a contact diagram.}
\centering
\begin{tabular}{|c|c|c|c|} 
\hline 
Process & s-cha & t-cha & Partial wave ($a_0$) \\
 \hline\hline
 $h V^1 \rightarrow h V^1$ & $V^1$ & $H,V^1$ & $\displaystyle-\frac{\lamL s}{64\pi\Mv[1]^2}\left(1 + 2 \frac{\lamL}{g^2}\frac{M_W^2}{\Mv[1]^2}\right) $ \\ \hline
 $h V^2 \rightarrow h V^2$ & $V^2$ & $H,V^2$ & $\displaystyle-\frac{\lambda_R s}{64\pi\Mv[2]^2}\left(1 + 2 \frac{\lambda_R}{g^2}\frac{M_W^2}{\Mv[2]^2}\right) $ \\ \hline
 $h V^\pm \rightarrow h V^\pm$ & $V^\pm$ & $H,V^\pm$ & $\displaystyle-\frac{\ld[2] s}{64\pi\Mvc^2}\left(1 + 2 \frac{\ld[2]}{g^2}\frac{M_W^2}{\Mvc^2}\right) $  \\ 
 \hline
\end{tabular}
\label{tab_partial}
\end{table}

\bibliography{bibliography}
\bibliographystyle{utphys}

\end{document}